\documentclass[preprint,5p,times, twocolumn]{elsarticle}

\usepackage{amssymb}
\usepackage{amsmath}
\usepackage[margin=0pt,font=small,labelfont=bf,labelsep=colon]{caption}
\usepackage{algorithm}
\usepackage{algpseudocode}
\usepackage{listings}
\usepackage{color}
\usepackage{textcomp}
\usepackage{url}
\definecolor{listinggray}{gray}{0.9}
\definecolor{lbcolor}{rgb}{0.9,0.9,0.9}
 \biboptions{sort&compress}

\journal{Computer Physics Communications}

\begin{document}

\begin{frontmatter}

\author{Zheyong Fan\corref{cor1}}
\ead{zheyong.fan@aalto.fi}
\cortext[cor1]{Corresponding author}
\author{Topi Siro}
\author{Ari Harju}

\title{Accelerated molecular dynamics force evaluation
on graphics processing units
for thermal conductivity calculations}

\address{COMP Centre of Excellence and Helsinki Institute of Physics,
Department of Applied Physics, Aalto University, Helsinki, Finland}

\begin{abstract}
Thermal conductivity is a very important property of a material 
both for testing theoretical
models and for designing better practical devices such as more efficient thermoelectrics,
and accurate and efficient calculations of thermal conductivity are very desirable.
In this paper, we develop a highly efficient molecular dynamics code fully
implemented on graphics processing units for thermal 
conductivity calculations using the Green-Kubo formula.
We compare two different schemes for force evaluation, a previously used
thread-scheme, where a single thread is used for one particle 
and each thread calculates the total force for
the corresponding particle, and a new block-scheme, where a whole block
is used for one particle and each thread in the block calculates one
or several pair forces between the particle associated with the given block
and its neighbor particle(s) associated with the given thread. For both schemes,
two different classical potentials, namely, the Lennard-Jones potential
and the rigid-ion potential are implemented. 
While the thread-scheme performs
a little better for relatively large systems, the block-scheme performs much better
for relatively small systems. The relative performance of the block-scheme over
the thread-scheme also increases with the increasing of the cutoff radius. 
We validate the implementation by calculating lattice thermal
conductivities of solid argon and lead telluride. The efficiency
of our code makes it very promising to study thermal conductivity
properties of more complicated materials, especially, materials
with  interesting nanostructures. 
\end{abstract}


\end{frontmatter}

\section{Introduction}
\label{Introduction}

Recently, graphics processing units (GPU) have attracted more and more attention
in the computational physics community due to their large-scale parallelism
ability, which has been explored in many kinds of computational physics
problems, including gravitational $N$-body simulation \cite{belleman2008},
classical molecular dynamics simulation \cite{yang2007,stone2007,vanmeel2008,
anderson2008,liu2008,friedrichs2009,rapaport2011},
classical Monte Carlo simulation \cite{preis2009},
quantum Monte Carlo simulation \cite{anderson2007},
quantum transport \cite{ihnatsenka2012}
and exact diagonalization \cite{siro2012}, just to name a few.
Despite the achievements obtained so far,
the potential of GPU to speed up
computational physics applications is far from being fully explored.

One of the physical problems which is very demanding of high performance
computing is molecular dynamics (MD) simulation and its GPU 
acceleration has been studied
by several groups \cite{yang2007,stone2007,vanmeel2008,
anderson2008,liu2008,friedrichs2009,rapaport2011}.
As early as in 2006, Yang \textit{et al} \cite{yang2007}
carried out a proof-of-principle
study on the acceleration of equilibrium molecular dynamics simulation
for thermal conductivity prediction of solid argon
based on the Green-Kubo formula using a GPU.
For this purpose, there is no need to update the neighbor
list since every particle in the system just
oscillates around its equilibrate
position. Using a very old GPU, they obtained
a speedup of about 10 relative
to a single CPU, although their implementation suffers from
accuracy loss, which results in a heat current autocorrelation function (HCACF) not
decaying to zero but a finite value.
After this pioneering work, a few other groups studied the GPU-acceleration
of more general MD simulations, mainly oriented to large-scale
simulations in biochemical systems, and the problem of neighbor list
update attracts special considerations
\cite{vanmeel2008,anderson2008,liu2008,rapaport2011}.
The reported speedups over CPU implementations
range from one to two orders of magnitude, depending on the interaction 
potential, the simulation size and the cutoff radius used in the simulations.

In this work, we consider the development and optimization of a full GPU
implementation of equilibrium MD simulations and use it
for thermal conductivity predictions using the Green-Kubo formula.
The main difference between our work and the previous ones
\cite{stone2007,vanmeel2008,anderson2008,liu2008,friedrichs2009,rapaport2011}
is that we are more interested in relatively small systems,
since size effects for
the Green-Kubo method of thermal conductivity prediction are not very
significant, and they can often be eliminated
with less than several thousand particles.
The demand for high performance computing for the Green-Kubo method
comes from the fact that the number of
simulation steps usually needs to be very large
(sometimes as large as $10^8$ steps \cite{yao2005})
to ensure a well converged HCACF
and the corresponding running thermal conductivity (RTC).
Around this simulation size,
we find that the conventional approach of force evaluation,
where a single thread is assigned to evaluate
the total force on one particle \cite{anderson2008},
can not utilize the computational power of modern GPUs sufficiently.
Here, we propose another approach of force evaluation,
where one block of threads is
assigned to evaluate the total force on one particle, which exhibits significantly
superior performance for small-to-intermediate simulation sizes.
In the one thread per particle scheme (which will be called the
thread-scheme), the number of invoked blocks
in the force evaluation kernel, which equals
the number of particles divided by the block size, is much smaller than
that in the one block per particle scheme
(which will be called the block-scheme),
where it equals the number of particles.
We will demonstrate that the block-scheme can attain its optimal
performance for relatively small systems and is several times faster than 
the thread-scheme for these.
Compared with the proof-of-principle
study of Yang \textit{et al} \cite{yang2007},
which only calculates the lattice thermal conductivity of solid argon using
the simple Lennard-Jones (LJ) potential, we also consider the
rigid-ion (RI) potential, which is more computational demanding.
On average, two orders of magnitude 
speedups can be obtained for the evolution part of the MD simulation.
For all the tested situations, the acceleration rates for the RI potential 
are several times larger than those for the LJ potential, reflecting the
higher arithmetic intensity for the RI potential.
In addition, neighbor list construction and HCACF
calculation are also implemented
in the GPU and good speedups are obtained for these parts too.

This paper is organized as follows.
In Sect. \ref{section:Backgrounds}, we review the techniques of
molecular dynamics simulation for thermal conductivity prediction using
the Green-Kubo formula and some important features of CUDA relevant to
our implementation. In Sect. \ref{section:Implementation}, we present
the detailed algorithms and implementations of our code.
In Sect. \ref{section:Performance}, the performances for the different
force evaluation schemes and the different
potentials are shown and compared.
In Sect. \ref{section:Validation}, we validate our implementation
by computing the lattice thermal conductivities of solid argon and
lead telluride at different temperatures. 
Our conclusions will be presented in Sect. \ref{section:Conclusion}.

\section{Background}
\label{section:Backgrounds}

\subsection{Green-Kubo method for thermal conductivity calculations}

In atomistic calculations of lattice thermal conductivity, both the
equilibrium-based and the non-equilibrium-based molecular dynamics
simulations can be employed. The non-equilibrium-based method uses
the Fourier's law of heat conduction and is referred to as the direct
method. The equilibrium-based method uses the
Green-Kubo linear response theory and is referred to as the
Green-Kubo method. Schelling \textit{et al} \cite{schelling2002}
showed that the equilibrium and the nonequilibrium
methods give consistent results. In some cases, the former
is more advantageous; in other cases, the reverse is true.
The most important advantage of the Green-Kubo method is that it has a much less
prominent size effect compared to the direct method.
In the direct method, unless the simulation cell is
many times longer than the mean-free path, scattering from the heat source
and heat sink contributes more to the thermal resistivity than the
intrinsic anharmonic phonon-phonon scattering does. In most cases,
a value for the thermal conductivity of an infinite system can be reliably
obtained by the direct method from simulations of different system sizes
and an extrapolation to an
infinite system size  \cite{schelling2002}.
However, as pointed out by Sellan \textit{et al} \cite{sellan2010},
the linear extrapolation procedure is only accurate when the
minimum system size used in the direct method simulations
is comparable to the largest mean free paths of the
phonons that dominate the thermal transport.
As for the computational cost, the Green-Kubo method requires
a rather long simulation time ($10^7 \sim 10^8$ time steps)
to get a well converged HCACF and thermal conductivity value.
To get a well defined temperature gradient, the direct method only requires
a relatively short simulation time ($\sim 10^6$ time steps).
However, the Green-Kubo method is still generally computationally cheaper
since one needs to simulate several very large systems when using the
direct method to accurately extrapolate to the
infinite limit. Additionally, by a single simulation, one can
obtain the full thermal conductivity
tensor of the system when using the Green-Kubo method, but can only obtain
a single component of the thermal conductivity when using the direct method.

According to the Green-Kubo linear response theory
\cite{green1954, kubo1957,mcquarrie2000},
the lattice thermal conductivity tensor
$\kappa_{\mu\nu}$ can be expressed as a time integral of
the HCACF $C_{\mu\nu}(t)$,
\begin{equation}
  \kappa_{\mu\nu} = \frac{1}{k_B T^2 V} \int_0^{\infty} dt C_{\mu\nu}(t),
\end{equation}
where $t$ is the correlation time, $k_B$ the Boltzmann's constant,
$V$ the volume and $T$ the temperature. The HCACF
$C_{\mu\nu}(t) = \langle J_{\mu}(0) J_{\nu}(t) \rangle$
is computed in the MD simulation by an average over
different time origins,
\begin{equation}
\label{equation:hcacf}
 C_{\mu\nu}(t) = \frac{1}{M} \sum_{m = 0}^{M-1}
 J_{\mu}(m \delta t) J_{\nu}(m \delta t + t),
\end{equation}
where $\delta t$ is the time step and $M$ is the number of time origins
to be averaged, which is approximately the number of production steps
if the number of correlation steps is much less than the number of production
steps.  For isotropic 3D materials, the thermal
conductivity scalar is usually defined to be the average of the
diagonal elements, $(\kappa_{xx} + \kappa_{yy} + \kappa_{zz}) / 3$.
Similarly, for a material isotropic in a plane, the thermal
conductivity in that plane can be defined to be
$(\kappa_{xx} + \kappa_{yy}) / 2$. Generally, the Green-Kubo method
is able to produce the full tensor of thermal conductivity in a single run.
We will use $\mathbf{C}$ to denote the vector with components
$C_{xx}$,  $C_{yy}$ and $C_{zz}$.

The heat current vector $\mathbf{J}$
(with components $J_{x}$, $J_{y}$ and $J_{z}$)
is defined to be the time derivative of the sum of the
moments of the site energies of the particles in the
system \cite{mcquarrie2000},
\begin{equation}
  \mathbf{J} = \frac{d}{d t} \sum_i^N \mathbf{r}_i E_i
  = \sum_i^N \mathbf{v}_i E_i + \sum_i^N \mathbf{r}_i \frac{d}{d t} E_i,
\end{equation}
where the site energy is the sum of the kinetic
energy and the potential energy,
$E_i = \frac{1}{2} m_i \mathbf{v}_i^2 + U_i$.
For a two-body potential, $U_i = \frac{1}{2} \sum_j U_{ij}$ and
$\mathbf{F}_{i} = \sum_i \mathbf{F}_{ij} = \sum_i
\left(-\frac{\partial}{\partial \mathbf{r}_i}\right) U_{ij} $,
we have
\begin{equation}
  \mathbf{J} = \sum_i^N \mathbf{J}_{i}
             = \sum_i^N
               \left[
               \sum_{j \neq i}^N \mathbf{J}_{ij}
               + \left( \frac{1}{2}m_i \mathbf{v}_i^2 \right) \mathbf{v}_i
               \right],
\end{equation}
where
\begin{equation}
\mathbf{J}_{ij} = \frac{1}{4}
    \left[
    \mathbf{v}_{ij} U_{ij}
  - \mathbf{r}_{ij} \left(\mathbf{F}_{ij} \cdot \mathbf{v}_{ij}\right)
    \right],
\end{equation}
with $\mathbf{r}_{ij} = \mathbf{r}_{j} - \mathbf{r}_{i}$
and $\mathbf{v}_{ij} = \mathbf{v}_{j} + \mathbf{v}_{i}$.

In this paper, we consider two kinds of pair-wise potentials.
The first is the LJ potential of the form
\begin{equation}
  U_{ij}^{\text{LJ}} = 4 \epsilon
  \left(
  \frac{\sigma^{12}}{r_{ij}^{12}} - \frac{\sigma^{6}}{r_{ij}^{6}}
  \right),
\end{equation}
where $\epsilon$ and $\sigma$ are two parameters of the model, the dimensions of
which are [energy] and [length] respectively. For argon,
$\epsilon/k_B = 119.8$ K and $\sigma = 3.405$ {\AA}, where $k_B$ is the
Boltzmann's constant. This potential has been used by many groups
to calculate thermal conductivities
of both pure solid argon
\cite{kaburaki1999,mcgaughey2004,tretiakova2004,chen2004,kaburaki2007}
and argon-krypton composites \cite{chen2005,landry2008}.
The second potential we consider is the RI potential
which consists of the Coulomb potential as well as a short range
part in the Buckingham form
\begin{equation}
  U_{ij}^{\text{Buckingham}}
  = A_{ij} \exp \left(-\frac{r_{ij}}{\rho_{ij}} \right)
  - \frac{C_{ij}}{r_{ij}^6}.
\end{equation}
The presence of the subscripts in the parameters
$A_{ij}$, $\rho_{ij}$ and $C_{ij}$ indicates that the values of
the parameters depend on
the particle types of the interacting pairs.

The other part of the RI potential is the Coulomb potential which is a long
range potential, and direct evaluation of it
using the traditional Ewald summation
is very time consuming for large systems.
Instead, the Wolf method \cite{wolf1999} is more
advantageous both computationally and conceptually.
In our work, we use a modified form of the Wolf formula
developed by Fennell \textit{et al} \cite{fennell2006}, 
in which both the potential and force are continuous at the cutoff radius,
\begin{equation}
\begin{split}
U_{ij}^{\text{Coulomb}}
 &=   \frac{q_{i}q_{j}}{4 \pi \epsilon_0}
     \left[ \frac{\text{erfc}(\alpha r_{ij})}{r_{ij}}
   - \frac{\text{erfc}(\alpha R_c)}{R_c}
\right. \\
&\left.
 +  \left(\frac{\text{erfc}(\alpha R_c)}{R_c^2}
 + \frac{2 \alpha}{\pi^{1/2}}
   \frac{\exp(-\alpha^2 R_c^2)}{R_c}\right) (r_{ij} - R_c) \right],
\end{split}
\end{equation}
where $\alpha$ and $R_c$ are the electrostatic damping factor and the cutoff
radius, respectively. The RI potential in the Buckingham
form is used extensively to
study lattice thermal conductivities of various kinds
of semiconductors and insulators such as
complex silica crystals \cite{mcgaughey2004b},
ZnO \cite{kulkarni2006} and PbTe \cite{qiu2011}.
In this work, we only consider PbTe, using the potential parameters
developed by Qiu \textit{et al} \cite{qiu2008}.
Note that in their parameterizations, partial charges are used for
both Pb and Te, which are +0.666 and -0.666, respectively.

To obtain the lattice thermal conductivity form HCACF, one usually
first fits the HCACF by a double-exponential function
\cite{che2000,mcgaughey2004}
with two time parameters and obtains the thermal conductivity
by analytical integration of the function. Ideally, the HCACF
should decay to zero except for some anomalous systems with
divergent thermal conductivity \cite{henry2008}.
Therefore, alternatively, one can also
directly integrate the HCACF and average the resulting RTC
values in an appropriate range
of time block \cite{schelling2002,sellan2010}.
We will show that the RTC will converge to a
definite value as long as the simulation time is large enough to
obtain a smooth HCACF which decays to zero after a given
correlation time.

\subsection{Overview of CUDA}

In this subsection, we review some techniques for GPU programming
with CUDA. We choose to use CUDA as our development tool, although
other tools such as OpenCL can be equally used. 
Only the important features relevant to our implementation
are presented here. For a more thorough introduction to CUDA,
we refer to the official manual \cite{cuda2011}. Although our
discussion is based on Tesla M2070, which is of compute capability 2.0,
the implementation and optimization can be easily ported to other platforms.

CUDA (Compute Unified Device Architecture) \cite{cuda2011}
is a parallel computing architecture for a hybrid platform
of CPU (the host) and GPU (the device).
Our CUDA program consists of both common C/C++ code which is
compiled and executed on the host and special functions
called kernels invoked from the host and executed on the
device. When a kernel is called from
the host, a grid of blocks, each of which contains individual threads, are
invoked to execute the instructions of the kernel in a
single instruction multiple data way. Each individual
thread has a unique ID which can be specified by built-in variables
such as \verb"threadIdx.x" and \verb"blockIdx.x".
The sizes of the grid and the blocks for a kernel
are specified at runtime
and can also be inferred from built-in variables
such as \verb"gridDim.x" and \verb"blockDim.x".

The concept of CUDA is closely connected to
the GPU hardware architecture, the knowledge
of which is vital for understanding and optimizing
CUDA applications. A GPU consists of a number of
streaming multiprocessors (SMs) and an SM
consists of a number of scalar processors (SPs).
For our testing GPU, Tesla M2070,
there are 16 SMs and $16\times32=448$ SPs.
The maximum numbers of resident blocks and threads that can be simultaneously
invoked within an SM are limited. For GPUs with compute capability 2.x,
they are 8 and 1536, respectively \cite{cuda2011}.
Thus, the theoretically
optimal number of threads in a block (the block size) is
about $1536/8=192$ for Tesla M2070.
In this case, the number of active blocks
and the number of active threads in an SM
all reach their optimized values.
In practice, the optimal block size is also affected by
the specific problem under consideration. For example,
in the block-scheme of force evaluation, a block
size of 128 is better, since 192 is not an exponential of 2,
and thus not suitable for binary reduction.

Understanding different GPU memories is also important for the implementation and
optimization of a CUDA program. There are various types of memories in the
GPU, each of which has its own strengths and limits.
The most important memories of the GPU are global memory, shared memory,
registers, constant memory and texture.
Global memory plays the crucial roles of
exchanging data between the CPU and the GPU and passing data from
one kernel to another.
Global memory can be both read and written by each thread in a kernel but
is very slow, thus should be minimized.
Shared memory plays the important role of sharing data within a block,
which is crucial when we do some reduction calculations such as summation.
Shared memory is fast but expensive. For Tesla M2070, there are only
48 KB shared memory for each SM. If we hope to keep
the maximum number  of resident blocks (which is 8 per SM),
we should not define more than 6 KB shared memory
in a kernel. Registers are private read-and-write memories
for individual threads.
The access of registers is very fast, but the amount of register memory
is rather limited. For Tesla M2070, there are only
32 K 32-bit registers for each SM. If we hope to keep
the maximum number of resident threads (which is 1024 per SM
if we take the block size as 128),
we should not define more than 32 32-bit registers
(or 16 64-bit registers) in a kernel.
Compared with global memory, constant memory is faster
but its amount is also very limited,
only 64 KB. Although some data in our application, such as
the neighbor list, do not change during the entire computation,
we cannot use constant memory for them since the required
amount of memory for the neighbor list exceeds the upper bound
even for a small system with a few hundred particles.
Alternatively, textures can be used for this kind of data.
We only use constant memory for
some invariant parameters needed in the kernels.

Lastly, we note that only when the number of blocks is several times larger
than the number of SMs to keep the GPU busy at all times, can we have
the possibility of attaining
the peak performance of the GPU. This important feature will be
discussed in detail when we compare the two schemes of force evaluation
in the following sections.

\section{Implementation details}
\label{section:Implementation}

In this section, we describe in some detail the techniques of the CUDA
implementation of our code.

\subsection{The overall consideration}

Firstly, we should find out which part of the CPU code deserves to be
rewritten using CUDA. Since force evaluation is the most time-consuming
part of an MD code, it is expected that a CUDA acceleration of this part
of the code would result in a significant speedup. However, this is not the
case due to the following two reasons: (1) even if the force evaluation
part takes up $99\%$ of the computation time, a 100-fold speedup of this
part would only result in about a 50-fold speedup for the whole program;
(2) if we only rewrite this part of code using CUDA, we have to exchange
data between CPU and GPU frequently, which is very time-consuming.
This will result in a lower speedup for the whole program.
Thus, we should implement the whole evolution part of the program in CUDA.

After confirming this, the next question is whether the whole evolution
part should be implemented in a single kernel or multiple kernels.
At first thought, the single kernel approach seems to be very appealing,
since in this way, global memory access can be reduced to its minimal
amount. Unfortunately, this is not easy to achieve. When we calculate
the total force exerted on one particle, we need the positions of all of
its neighbor particles, some of which would reside in different blocks
from the one where the considered particle resides in. Since there is no
intrinsic way to synchronize the individual blocks \cite{cuda2011},
the positions of the neighbor particles cannot be guaranteed
to be completely updated before evaluating the total force
of a given particle. Therefore, the natural choice is to
implement multiple kernels for the evolution part, some devoting to the
updating of positions and velocities and one devoting to the force
evaluation.

Thus, the overall structure of the whole CUDA program is as follows.
\begin{enumerate}
 \item Allocate global memory in the GPU according to the input parameters
       and transfer data from CPU to GPU.
 \item Do the simulation in the GPU.
     \begin{enumerate}
      \item Construct the invariant neighbor list.
      \item Evolve the system according to the interaction potential.
            In the equilibrium stage, control the
            temperature and / or pressure; in the production stage,
            record the heat current data for each time step into
            global memory.
     \item Calculate the HCACF from the recorded heat current data.
     \end{enumerate}
 \item Transfer the HCACF data from GPU to CPU and analyze the results.
\end{enumerate}

To facilitate later discussions, we list the main data in the global memory
of the GPU and the relevant parameters:
\begin{enumerate}
 \item $N$ is the number of particles in the simulated system.
 \item $N_c$ is the number of correlation steps.
 \item $N_p$ is the number of production steps.
 \item NN$_i$ ($0 \leq i \leq N-1$) is the number of neighbor particles for particle $i$.
 \item NL$_{ik}$ ($0 \leq i \leq N-1$, $0 \leq k \leq \text{NN}_i-1$)
       is the index of the $k$-th neighbor particle of particle $i$.
 \item $\mathbf{r}_i$, $\mathbf{v}_i$, $\mathbf{F}_i$ and
       $\mathbf{J}_i$ ($0 \leq i \leq N-1$) are the position,
       velocity, force and heat current for particle $i$ at a given time step.
 \item $\mathbf{J}(i)$ ($0 \leq i \leq N_p-1$) is the total heat current
       of the system at time step $i$.
 \item $\mathbf{C}(i)$ ($0 \leq i \leq N_c-1$) is the $i$-th HCACF data.
\end{enumerate}

We now discuss the details of the implementations of the relevant kernels
in the GPU.

\subsection{The neighbor list construction kernel}

For our purpose, the simple Verlet neighbor list \cite{verlet1967} suffices.
Although in our applications, the neighbor list needs no update during
the simulation, it is still advantageous to implement it directly in the GPU,
which can reduce the computation time significantly
compared with a direct $O(N^2)$ implementation in the CPU. We use a
simple algorithm to construct the Verlet neighbor list as listed in
\textbf{Algorithm \ref{algorithm:neighbor}}.
This kernel is launched with
the execution configuration \verb"<<<"$\lceil N/S_b\rceil$, $S_b$\verb">>>",
where $N$ is the number of particles
and $S_b$ is the block size. When $N$ is not an integer multiple of
$S_b$, the indices of some threads in the kernel would exceed $N$.
The \textbf{if} statement in line 2 ensures that only
the valid memory is manipulated. After executing this kernel,
the number of neighbor particles for particle $i$ is stored
in NN$_i$ and the index of the $k$-th neighbor particle of particle $i$
is stored in NL$_{ik}$.

Note that we have expressed the neighbor list in a matrix from $N_{ik}$,
but the actual order of the data in this matrix still needs to be specified when
we implement it in the computer code.
There are two natural ways to arrange the data. The first is to
store all the indices of the neighbor particles of the $i$-th
particle consecutively, i.e., in the order of
NL$_{00}$, NL$_{01}$, NL$_{02}$, $\cdots$,
NL$_{10}$, NL$_{11}$, NL$_{12}$, $\cdots$,
NL$_{i0}$, NL$_{i1}$, NL$_{i2}$, $\cdots$.
The second is to store the indices of the $k$-th neighbor particles
for all the particles consecutively, i.e., in the order of
NL$_{00}$, NL$_{10}$, NL$_{20}$, $\cdots$,
NL$_{01}$, NL$_{11}$, NL$_{21}$, $\cdots$,
NL$_{0k}$, NL$_{1k}$, NL$_{2k}$, $\cdots$.
While for a serial CPU implementation, the first choice is more
preferable, it turns out that, for our GPU implementation, different
force evaluation schemes require different forms of the neighbor list
to ensure coalesced global memory access. This will be discussed in more
detail when we present the force evaluation algorithms.

\begin{algorithm}
\caption{Pseudo code for Verlet neighbor list construction kernel.}
\label{algorithm:neighbor}
\begin{algorithmic}[1]
\Require $b$ is the block index
\Require $t$ is the thread index
\Require $S_b$ is the block size
\Require $i = S_b\times b+t$
\State $k \leftarrow 0$
\If {$i<N$}
 \For {$j$ = 0 to $N - 1$}
  \If {$j \neq i$}
   \State $\mathbf{r}_{ij} \leftarrow$
          minimum image of $\mathbf{r}_{j} - \mathbf{r}_{i}$
   \If {$|\mathbf{r}_{ij}|^2 <$  square of the cutoff radius}
    \State NL$_{ik} \leftarrow j$
    \State $k \leftarrow k + 1$
   \EndIf
  \EndIf
 \EndFor
 \State  NN$_i \leftarrow k$
\EndIf
 \end{algorithmic}
 \end{algorithm}

\subsection{The integration kernels}

The update of positions and velocities for one particle is independent
of that for another particle. Therefore, they can be carried out in
parallel. Since this operation is rather simple, we can assign the task
of updating the velocity and position of a particle to a single thread.
Thus, the execution configuration of any of the integration kernels is the same
as that of the neighbor list construction kernel. Again,
an \textbf{if} statement is
necessary to ensure that invalid memory is not manipulated.
The algorithms for these kernels are rather straightforward and
not provided here.

\subsection{The Force evaluation kernel}

Force evaluation is the most time consuming part of most MD simulations and
thus deserves special consideration.

\subsubsection{The thread-scheme for force evaluation}

Since the calculation of the total force
for one particle is independent of the calculation of the total force for
any other particles, the most natural choice for GPU implementation of the
force evaluation kernel is to assign a single thread to one particle and
loop for all the neighbor particles of it to accumulate the total
force exerted on this particle. To our knowledge, previous works
either follow more or less this strategy when using the neighbor list approach
or use a cell-list approach for force evaluation instead.
The pseudo code for this
thread-scheme is presented in
\textbf{Algorithm \ref{algorithm:force_thread}}.
The total force and heat current for one particle are
accumulated (lines 10 and 11) in the \textbf{for} loop and saved (line 13)
to global memory after exiting the \textbf{for} loop.

Global memory access is relatively slow compared with the access of other
memories such as shared memory and registers in the GPU,
and we have made many efforts to minimize global memory access as much
as we can, although it is not manifest from the presented pseudo codes.
For example, in \textbf {Algorithm  \ref{algorithm:force_thread}},
we have to calculate the distances between one particle
and all its neighbor particles $\mathbf{r}_{ij}$ ($0\leq j<\text{NN}_i$).
While we need to read in the positions $\mathbf{r}_j$ ($0\leq j<\text{NN}_i$)
from the global memory within the \textbf{for} loop,
we need not read in the position $\mathbf{r}_i$ repeatedly within
the \textbf{for} loop. Instead, we can read in $\mathbf{r}_i$ before entering
the \textbf{for} loop and store it in a register. As pointed out earlier,
registers are fast but their number for each SM is limited, and
excessive use of them will deteriorate the performance. In the case that registers
are not enough for use, we can use some shared memory
substituting for registers.

When a global memory access is unavoidable,
it is important to maximize the coalescing
by using the most optimal access pattern possible.
Generally, the positions $\mathbf{r}_j$ and velocities $\mathbf{v}_j$
within the \textbf{for} loop are accessed in a random pattern, and
a special particle sorting method has been developed to generate better
memory access pattern \cite{anderson2008}. As for the global memory
access for the neighbor list as given in line 5, we note that here
we should choose the neighbor list representation where
the indices of the $k$-th neighbor particles
for all the particles are stored consecutively.
This makes nearby threads in a warp access nearby elements
in the neighbor list.

In the thread-scheme, the execution configuration of the
force evaluation kernel is the same as that of the neighbor list
construction kernel, for which the number of
blocks invoked is $\lceil N/S_b \rceil$, which is not large enough
to fully utilize the computational resources of a modern GPU
for relatively small systems.
This motivates us to consider the other force evaluation scheme,
as discussed below.

\begin{algorithm}
\caption{Pseudo code for the force evaluation kernel in the thread-scheme.}
\label{algorithm:force_thread}
\begin{algorithmic}[1]
\Require $b$ is the block index
\Require $t$ is the thread index
\Require $S_b$ is the block size
\Require $i=S_b\times b+t$
\State $\mathbf{F}_{i} \leftarrow 0$
\State $\mathbf{J}_{i} \leftarrow 0$
\If {$i<N$}
 \For {$k$ = 0 to NN$_i - 1$}
   \State $j \leftarrow \text{NL}_{ik}$
   \State $\mathbf{r}_{ij} \leftarrow $ minimum image of
          $\mathbf{r}_{j} - \mathbf{r}_{i}$
   \State $\mathbf{v}_{ij} \leftarrow \mathbf{v}_i + \mathbf{v}_j$
   \State Calculate the pair force $\mathbf{F}_{ij}$
   \State Calculate the pair heat current $\mathbf{J}_{ij}$
   \State $\mathbf{F}_{i} \leftarrow \mathbf{F}_{i} + \mathbf{F}_{ij}$
   \State $\mathbf{J}_{i} \leftarrow \mathbf{J}_{i} + \mathbf{J}_{ij}$
 \EndFor
 \State Save $\mathbf{F}_{i}$ and $\mathbf{J}_{i}$ to global memory
\EndIf
 \end{algorithmic}
\end{algorithm}

\subsubsection{The block-scheme for force evaluation}

To increase the number of blocks in the force evaluation kernel, we notice
that for force evaluation, there is a second level of parallelism: the
calculations of the pair forces between one particle and all its neighbor
particles are also independent of each other. Therefore, instead of
accumulating the total force of a particle in a single thread, we can
also calculate the pair forces between the given particle and all its
neighbor particles in different threads. These threads should be in
the same block in order to accumulate the total force of the considered
particle in an efficient way by making use of shared memory. In the
simplest case, one thread in the block only computes a single pair force.
For example, (block $i$, thread $j$) is assigned to calculate the
force acted on particle $i$ from the $j$-th neighbor particle of $i$.
In this case, the number of blocks for the force evaluation kernel is
exactly the number of particles in the system, which is much larger
than the total number of SMs in the GPU for a system with intermediate
size. We call this the block-scheme for force evaluation and the
implementation is given in \textbf{Algorithm~\ref{algorithm:force_block}}.

As in the case of the thread-scheme, the choice of
the neighbor list representation is important.
Since in the block-scheme nearby threads in a warp
evaluate the pair forces between a given particle and
a portion of its neighbor particles, we should choose the
neighbor list representation where the indices of the neighbor
particles of the $i$-th particle are stored consecutively.

\begin{algorithm}
\caption{Pseudo code for the force evaluation kernel in the block-scheme.}
\label{algorithm:force_block}
\begin{algorithmic}[1]
\Require $i$ is the block index
\Require $k$ is the thread index
\Require $\mathbf{F}_{ij}$ and $\mathbf{J}_{ij}$ are in shared memory
\State $\mathbf{F}_{ik} \leftarrow 0$
\State $\mathbf{J}_{ik} \leftarrow 0$
    \If {$k < \text{NN}_{i}$}
       \State $j \leftarrow \text{NL}_{ik}$
       \State $\mathbf{r}_{ij} \leftarrow $
              minimum image of $\mathbf{r}_{j} - \mathbf{r}_{i}$
       \State $\mathbf{v}_{ij} \leftarrow \mathbf{v}_i + \mathbf{v}_j$
       \State Calculate the pair force $\mathbf{F}_{ij}$
       \State Calculate the pair heat current $\mathbf{J}_{ij}$
    \EndIf
\State synchronize the threads in each block
\State Binary reductions of $\mathbf{F}_{ij}$ and $\mathbf{J}_{ij}$
       to $\mathbf{F}_{i0}$ and $\mathbf{J}_{i0}$
\State Save $\mathbf{F}_{i0}$ and $\mathbf{J}_{i0}$ to global memory
       $\mathbf{F}_{i}$ and $\mathbf{J}_{i}$
\end{algorithmic}
\end{algorithm}

In the block-scheme, shared memory must be used for recording
the pair forces $\mathbf{F}_{ij}$
and the corresponding heat current components
$\mathbf{J}_{ij}$.
Otherwise, there is no efficient way to sum them up to get the
total force and heat current for a given particle. Before
recording the data, we should initialize all the elements of
$\mathbf{F}_{ij}$ and $\mathbf{J}_{ij}$ to be zero, as shown in
lines 1 and 2, since the number of neighbor particles
for some particles would probably be less than the block size.
After recording the calculated data into
$\mathbf{F}_{ij}$ and $\mathbf{J}_{ij}$,
we should synchronize the threads in each block before
summing them up. The synchronization ensures that all the
threads have completed their calculations before the
summation.  The summation can be carried out efficiently
in the way of binary reduction, i.e., add the second half
of data to the first half and repeat the process until
we get a single number, which is the sum of the whole data.

The execution configuration for the force evaluation kernel
in the block-scheme deserves more consideration. Firstly,
the block size should be an exponential of two, such as
128, in order to do the binary reductions for
$\mathbf{F}_{ij}$ and $\mathbf{J}_{ij}$.
Secondly, since the number of blocks in the kernel
equals the number of particles,
which has the chance of exceeding the maximal value of the
first dimension of the grid size, which is 65535 for
Tesla M2070, we need to use a two dimensional grid in such a
way that the difference between the total number blocks in the
grid and the number of particles is minimized.

In the above discussion of the block-scheme, we only considered
the simplest case, where a whole block of threads are devoted to
calculating the total force of a single particle and each thread
only calculates one pair force. In general, we can have two
alternatives, depending on the values of the block size $S_b$
and the maximal number of neighbor particles $N_m$.
If $N_m > S_b$, we should calculate at most $\lceil N_m / S_b \rceil$ pair
forces in each thread. For example, if $N_m = 256$ and
$S_b = 128$, each thread has to calculate two pair forces.
In this case, it is important to access the global memory in
a coalesced way. This requires to use thread $j$ to calculate
the pair forces associated with the $j$-th and $(j+S_b)$-th
neighbor particles, instead of those associated with two adjacent
neighbor particles. If $N_m \leq S_b / 2$, we do not need to use
a whole block for one particle. For example, if $N_m = 64$ and
$S_b = 128$, we can use one block for two particles, with the
first 64 threads for the first one and the second 64 threads
for the second one.

After the execution of the force evaluation kernel, either in the
thread-scheme or in the block-scheme, the total forces $\mathbf{F}_{i}$
and heat currents $\mathbf{J}_{i}$ for the individual particles
are stored into global memory, and we should, at each time step,
sum up the heat current values for the individual particles to get the heat
current for the whole system, $\mathbf{J} = \sum_i \mathbf{J}_{i}$.
This can also be done efficiently by using the binary reduction.

\subsection{HCACF calculation}

As discussed in the beginning of this section, to obtain a high acceleration
rate for the whole program, one has to implement the whole evolution part
in the GPU. Although the post-processing part of the program, namely, the
part for HCACF calculation usually takes up only a fraction of computation
time of the whole program, it is still advantageous to implement this part
on the GPU rather than on the CPU, since when we have obtained orders of magnitude of
speedup for the evolution part, the post-processing part would take up
a significant portion of the whole computation time.

The GPU implementation of HCACF calculation is very straightforward. Since
the ensemble averages (which are time averages in MD) of the HCACF
at different correlation times can be performed independently, we can
simply use one block for one point of the HCACF data.
The pseudo code
for the HCACF calculation kernel as presented in
\textbf{Algorithm}~\ref{algorithm:HCACF} is designed according to
Eq. (\ref{equation:hcacf}).
This kernel is executed with the configuration of
\verb"<<<"$N_c$, $S_b$\verb">>>", where $N_c$ is the
number of correlation steps
and $S_b$ is the block size. As in the case of
the force evaluation kernel in the block-scheme, if the number of
blocks (which is $N_c$ here) exceeds the upper bound of the
first dimension of the grid size (which is 65535 for Tesla M2070),
a two dimensional grid is needed.
After executing this kernel, the HCACF data are saved in global
memory, which will be transferred to CPU for analysis.

The value of $M$ in Eq. (\ref{equation:hcacf}), which is the number of time
origins used to calculate the ensemble averages of the HCACF data is
chosen to be $S_b \lfloor (N_p - N_c) / S_b \rfloor$
for any correlation time. This will waste a small portion of heat
current data if $N_p - N_c$ is not an integer multiple of $S_b$.
Usually, $N_p$ is much larger than both $N_c$ and $S_b$, and we have
$M \approx N_p$.

\begin{algorithm}
\caption{Pseudo code for HCACF calculation kernel.}
\label{algorithm:HCACF}
\begin{algorithmic}[1]
\Require $i$ is the block index
\Require $j$ is the thread index
\Require $S_b$ is the block size
\Require $\mathbf{C}_{ij}$ is in shared memory
\State $\mathbf{C}_{ij} \leftarrow 0$
\For {$n = 0$ to $\lfloor (N_p - N_c) / S_b \rfloor - 1$}
\State $\mathbf{C}_{ij} \leftarrow \mathbf{C}_{ij} +
        \mathbf{J}(j+nS_b) \mathbf{J}(j+nS_b+i)$
\EndFor
\State synchronize the threads in each block
\State binary reduction of $\mathbf{C}_{ij}$ to $\mathbf{C}_{i0}$
\State Save $\mathbf{C}_{i0}$ to global memory $\mathbf{C}(i)$
 \end{algorithmic}
 \end{algorithm}

\subsection{Use of texture memory}

Texture memory is a kind of read-only memory that is cached on-chip.
In some situations, it provides higher effective bandwidth,
especially in the cases that memory access patterns exhibit a great
deal of spatial locality, i.e., nearby threads are likely to read from
nearby memory locations. This is the case for the force evaluation
kernel (both the thread-scheme and the block-scheme), where the
access of the global memory containing the data of the positions $\mathbf{r}_i$
and velocities $\mathbf{v}_i$ of the neighbor particles for a given particle
is not coalesced but exhibits some spatial locality.
The overall gain of performance by using
texture memory is about $10\%$.

\subsection{Global memory requirements}

Lastly, we give an estimation of the memory requirements with respect
to the simulation size and simulation time. Most of the global memory
is occupied by the neighbor list and the heat current data. For the
neighbor list, the required global memory scales with the number of
particles $N$ as 2 $N/10^6$ GB if the maximum number of neighbor particles
for one particle is less than 512. For the heat current data, the required
global memory scales with the number of production steps $N_p$ as
1.2 $N_p/10^8$ GB and  2.4 $N_p/10^8$ GB for single and double precisions,
respectively. Thus, it is quite safe to perform a simulation with the
system size as large as $N=10^6$ and the number of production steps as
large as $N_p=10^8$ for Tesla M2070, which has 6 GB of global memory.
The Green-Kubo method rarely needs a simulation domain size as large
as one million particles.  Regarding the simulation
time, if $10^8$ heat current data are not enough to obtain a well
converged HCACF, we can simply perform the same simulation for a number of
times with different initial random velocities and average the resulting
HCACFs to get a better one.

\section{Performance measurements}
\label{section:Performance}

In this section, we measure the performance of our
GPU code running on a Tesla M2070 and
compare it with that of our CPU code running on an
Intel Xeon X5650.
The CPU code is implemented in C/C++ and is compiled using g++
with the O3 optimization mode. Newton's third law is 
also used to save unnecessary calculations for the CPU code.
However, as pointed out by others \cite{yang2007,anderson2008}, 
it is not straightforward and beneficial 
to use the Newton's third law in the GPU. 
Our GPU implementation thus does not use this law. 

\subsection{The evolution part}

\begin{figure*}
\begin{center}
\begin{tabular}{ccc}
  \includegraphics[width=2.2 in]{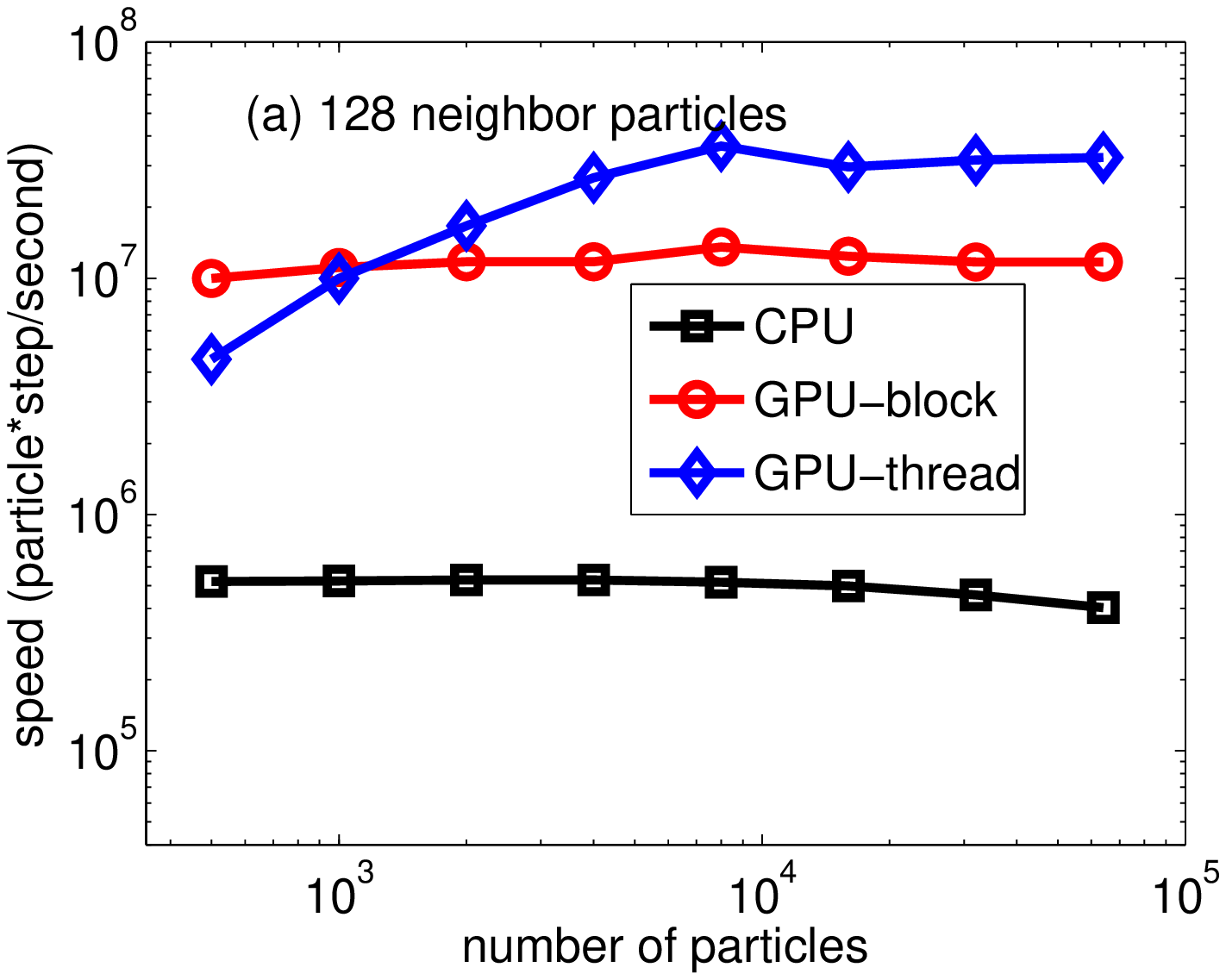}&
  \includegraphics[width=2.2 in]{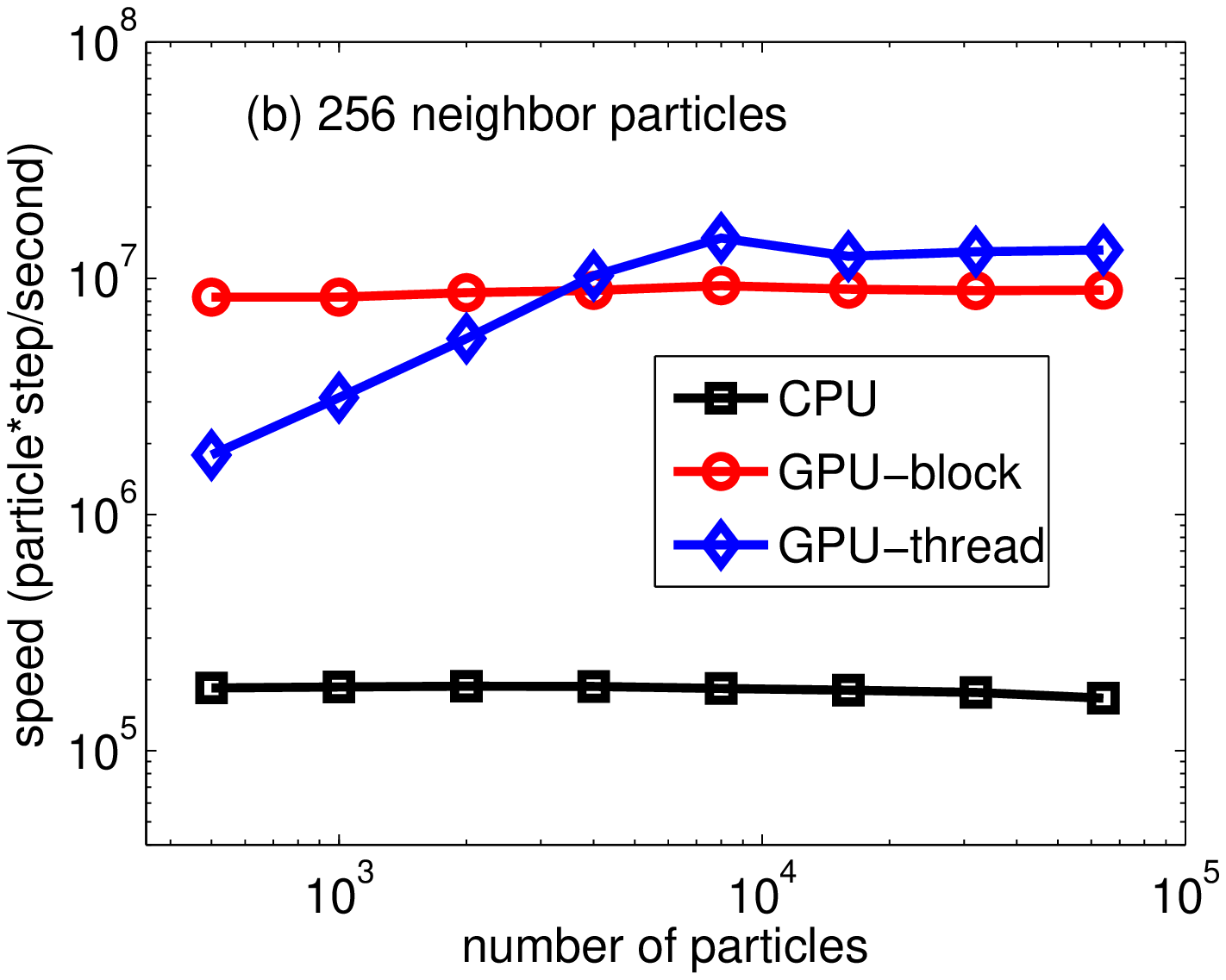}&
  \includegraphics[width=2.2 in]{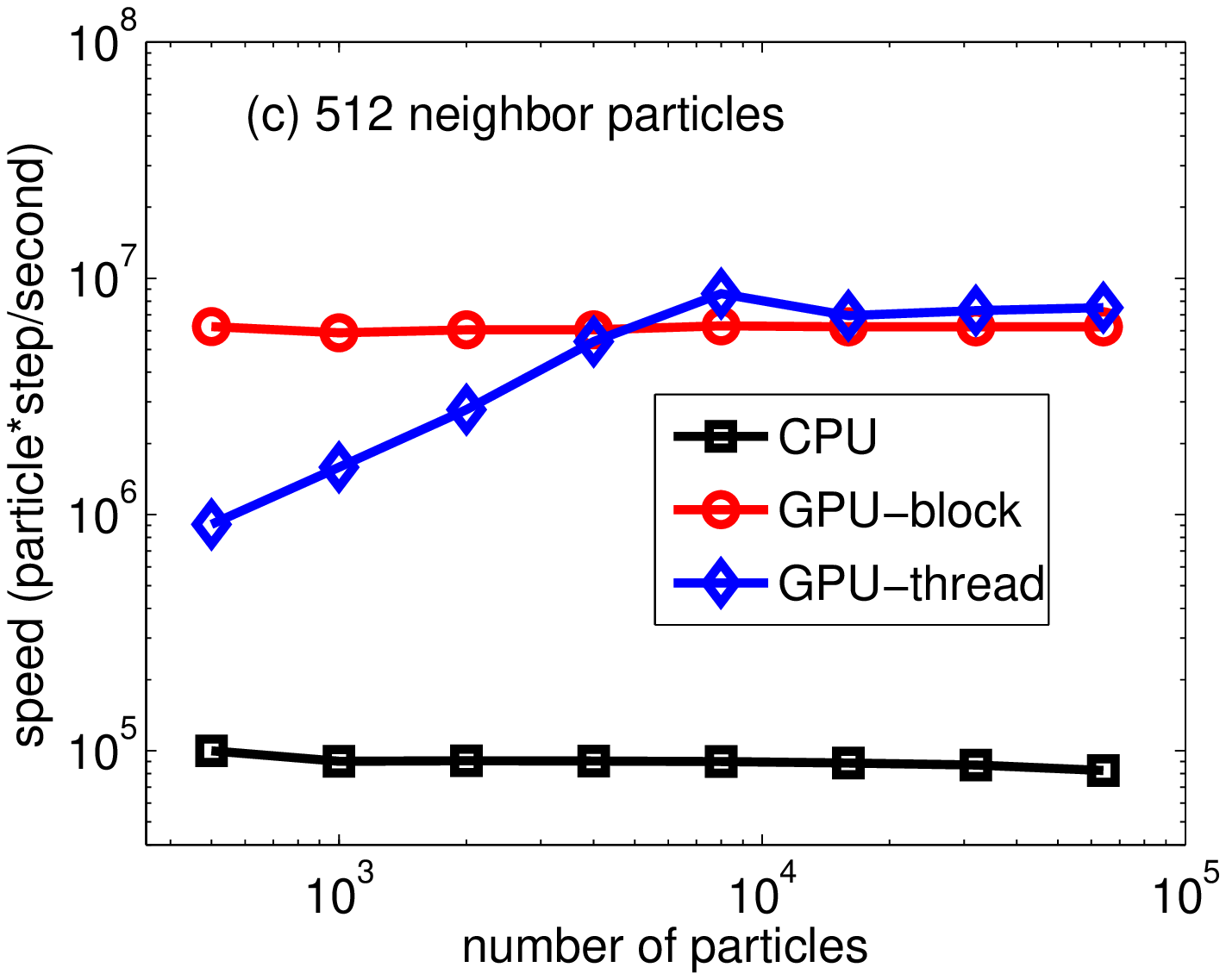}\\
  \includegraphics[width=2.2 in]{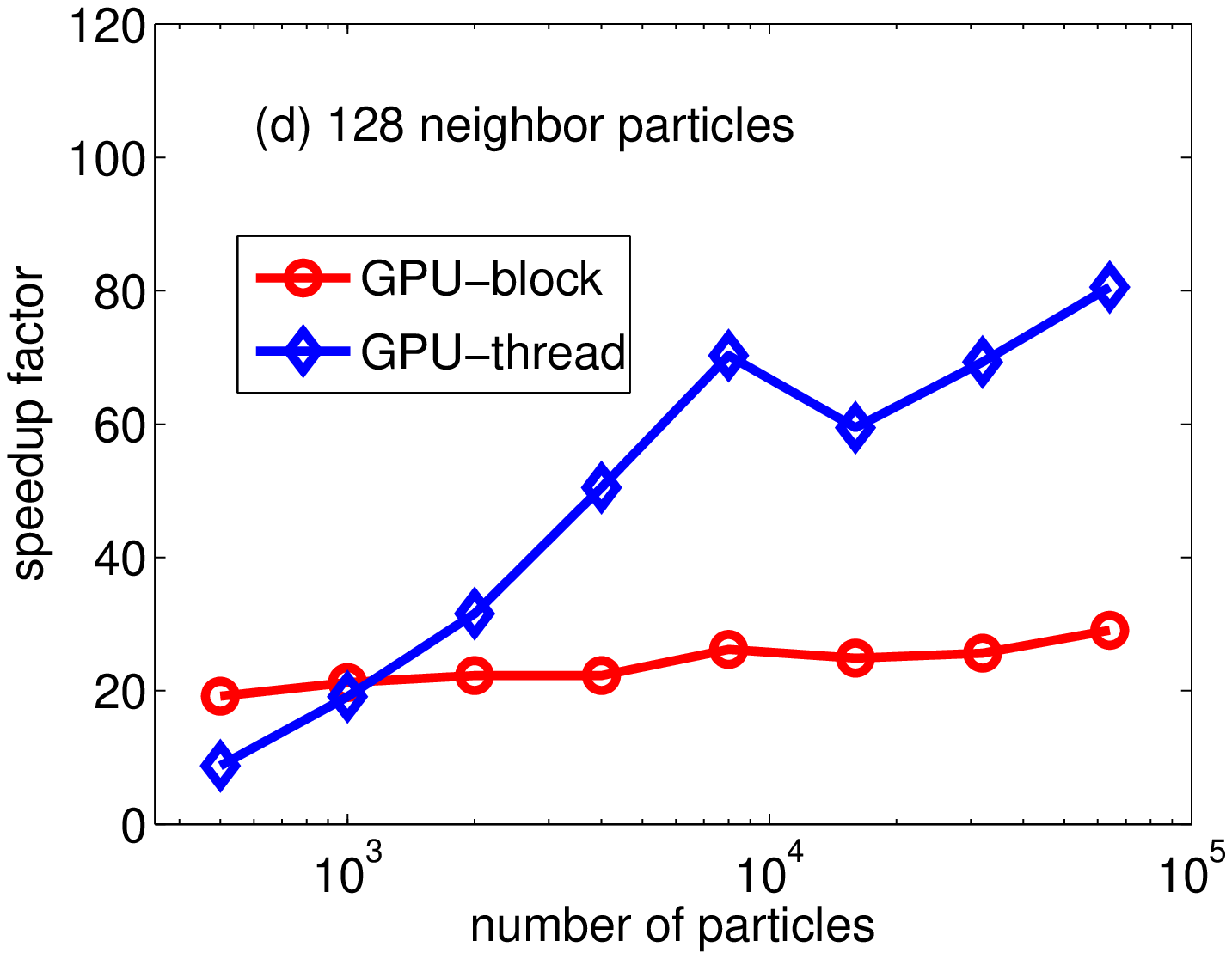}&
  \includegraphics[width=2.2 in]{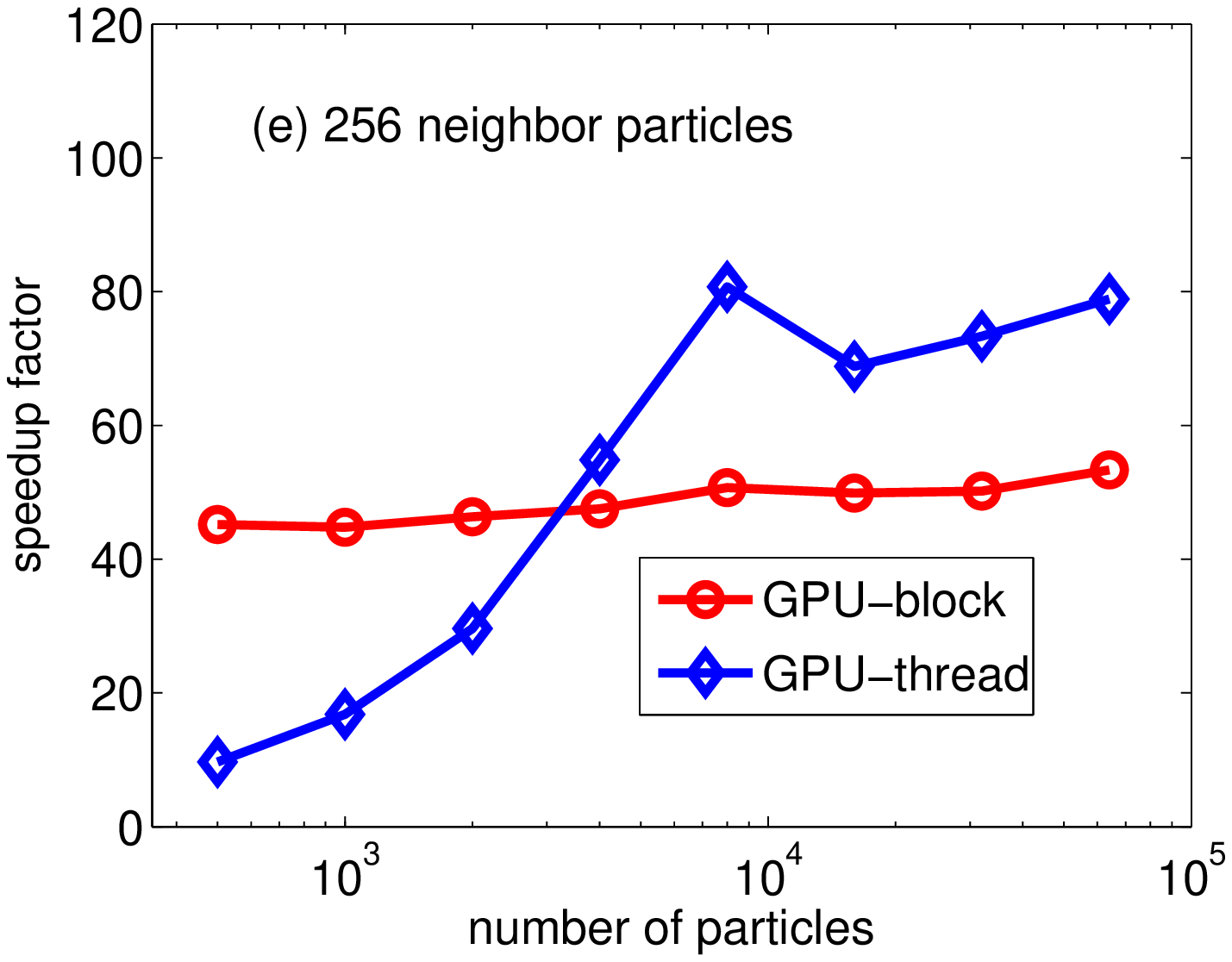}&
  \includegraphics[width=2.2 in]{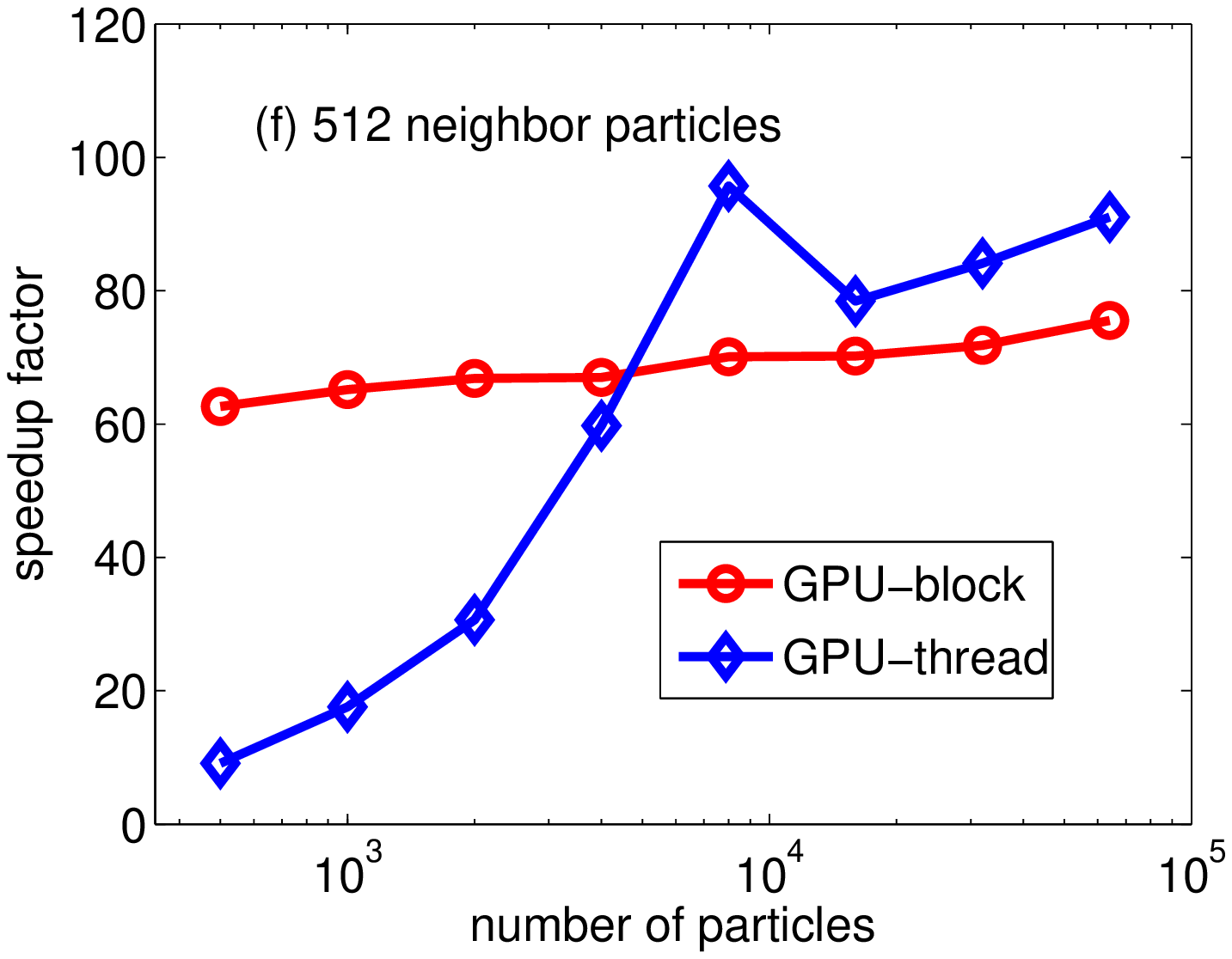}\\
  \includegraphics[width=2.2 in]{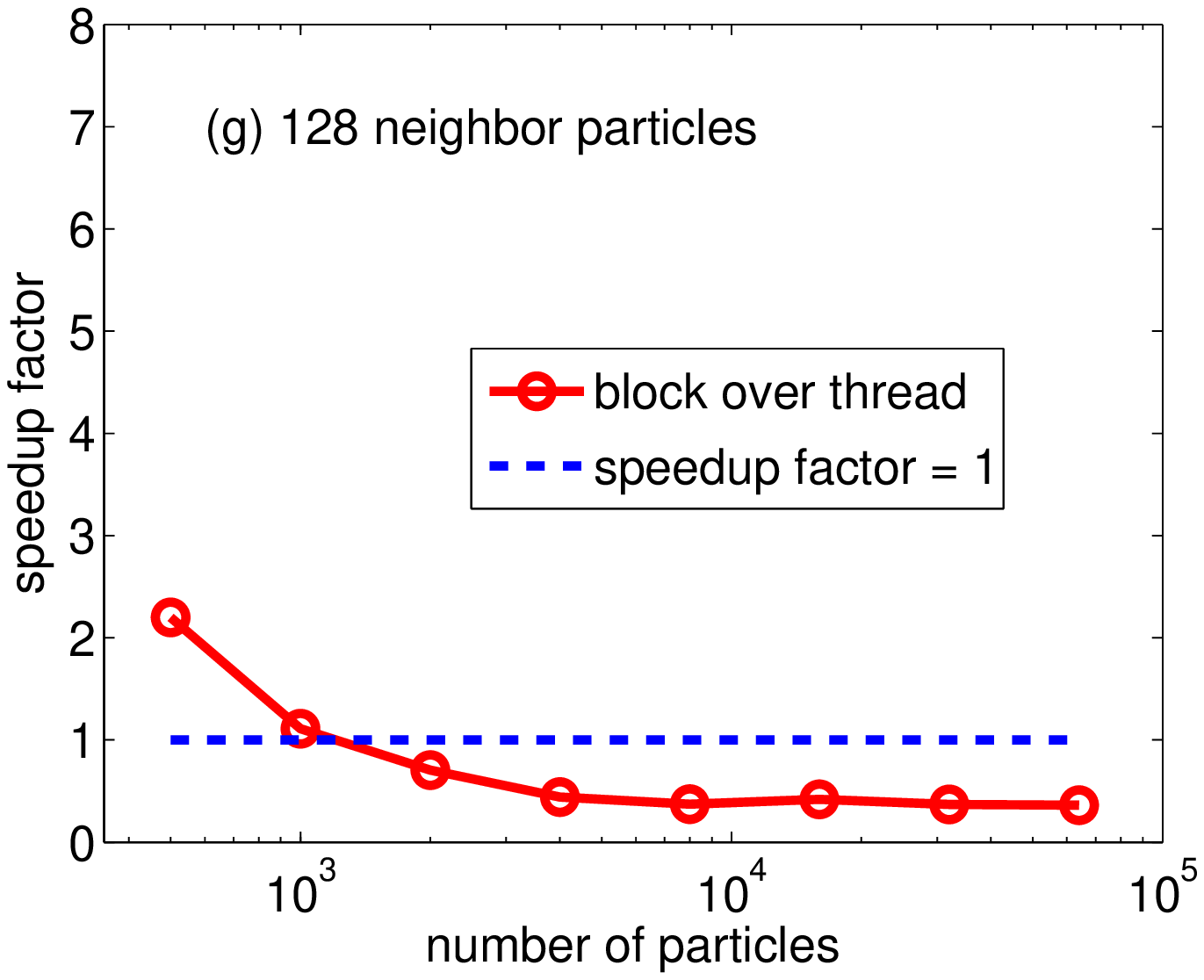}&
  \includegraphics[width=2.2 in]{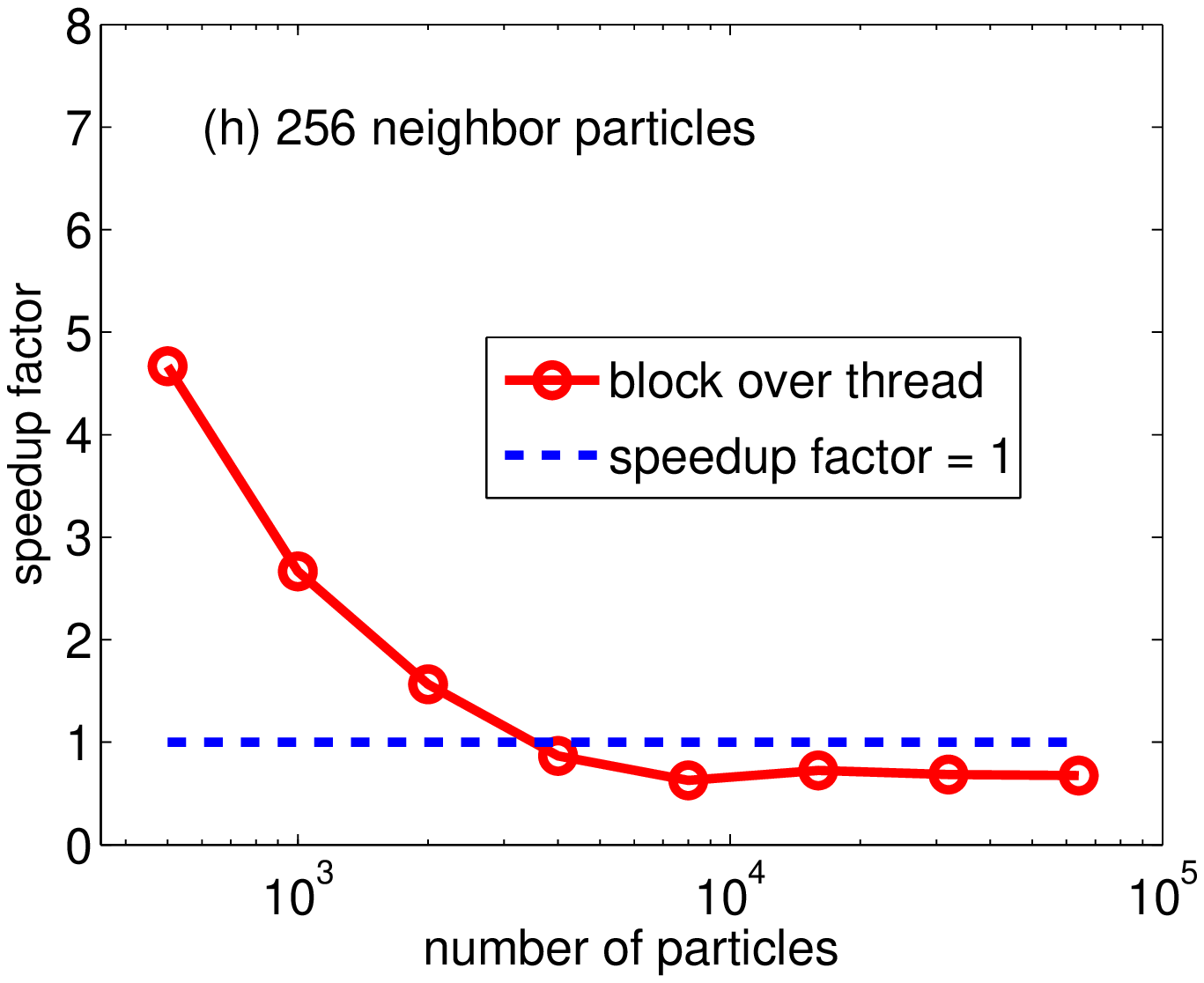}&
  \includegraphics[width=2.2 in]{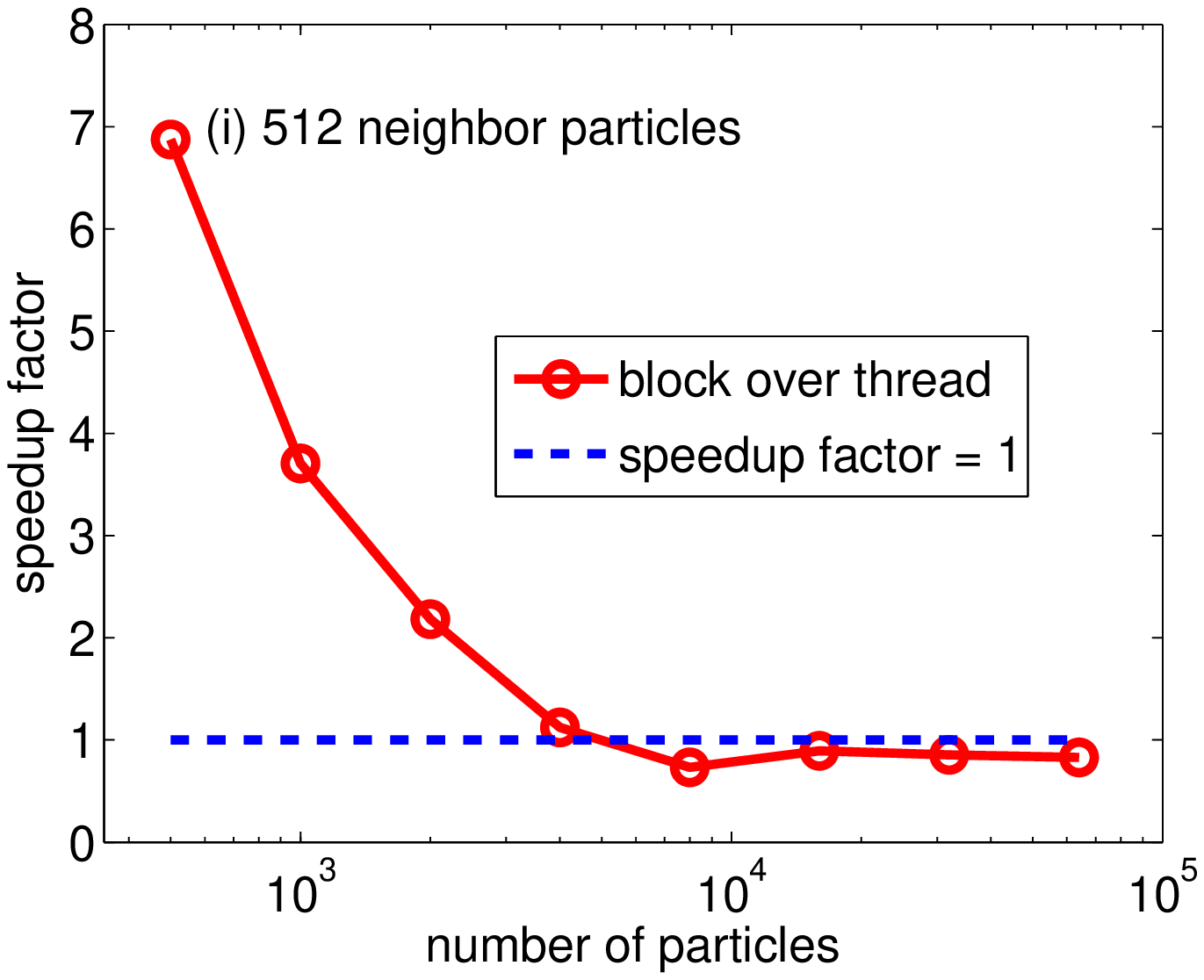}\\
\end{tabular}
  \caption{(Color online) Performance evaluation for the evolution part 
           in the production stage
           for the LJ potential: (a-c) computation speeds for both the CPU code 
           and the GPU code with different force evaluation schemes; (d-f) speedup
           factors of the GPU code over the CPU code; (g-i) relative speedup factors
           of the block-scheme over the thread-scheme.
           The maximal number of neighbor particles
           in the simulations (128, 256 or 512) 
           are indicated in each subplot.      
}
\label{figure:argon_performance}
\end{center}
\end{figure*}

\begin{figure*}
\begin{center}
\begin{tabular}{ccc}
  \includegraphics[width=2.2 in]{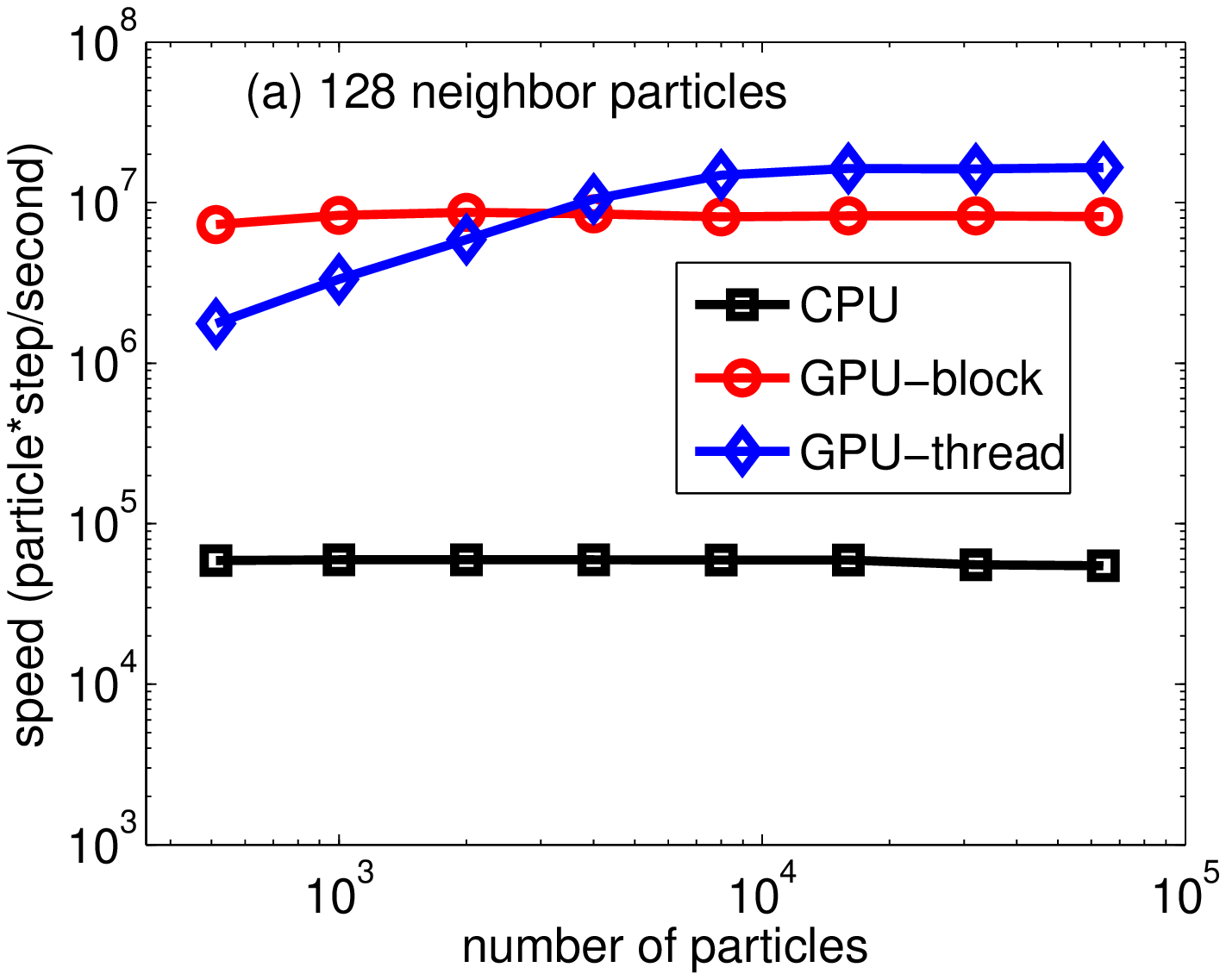}&
  \includegraphics[width=2.2 in]{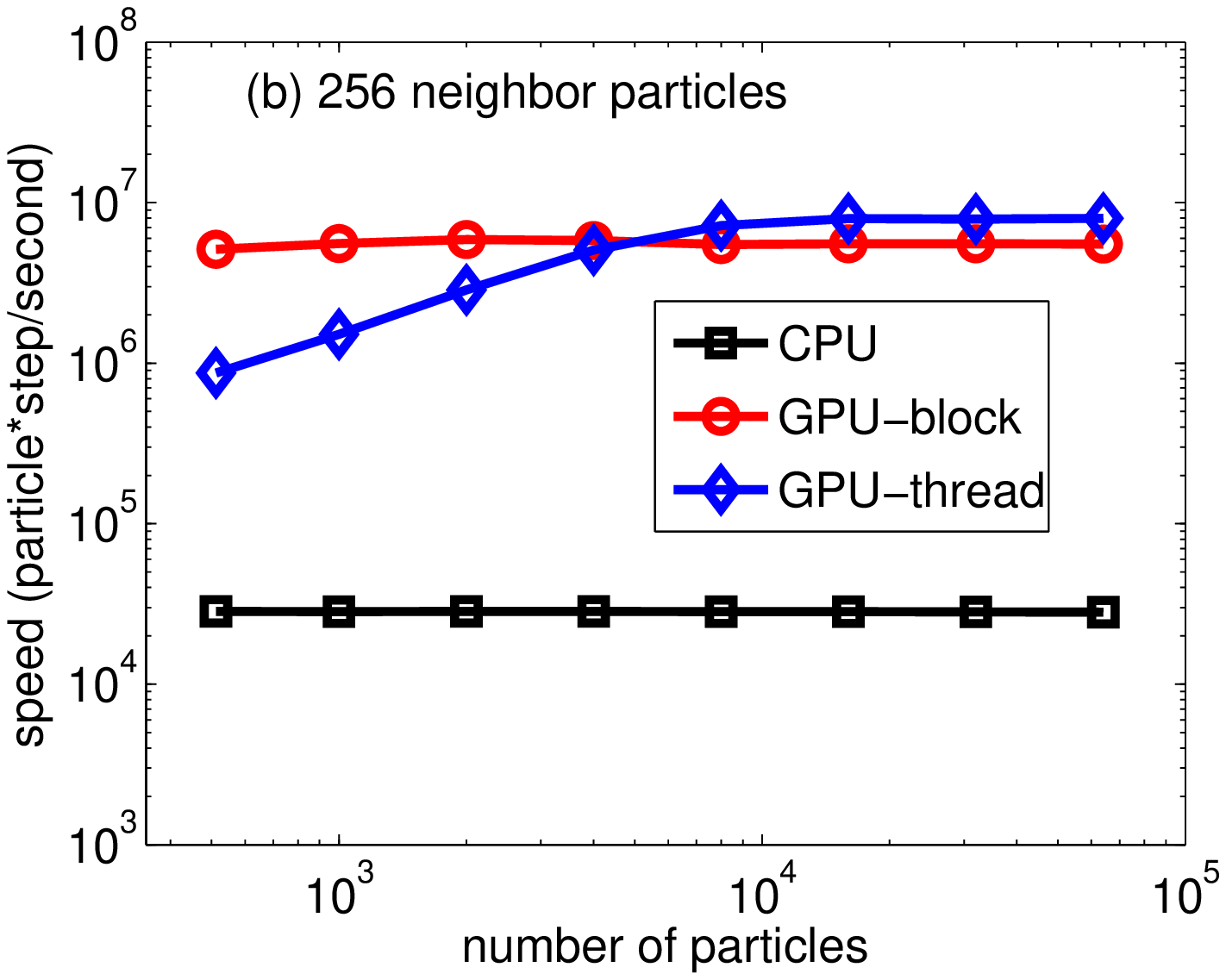}&
  \includegraphics[width=2.2 in]{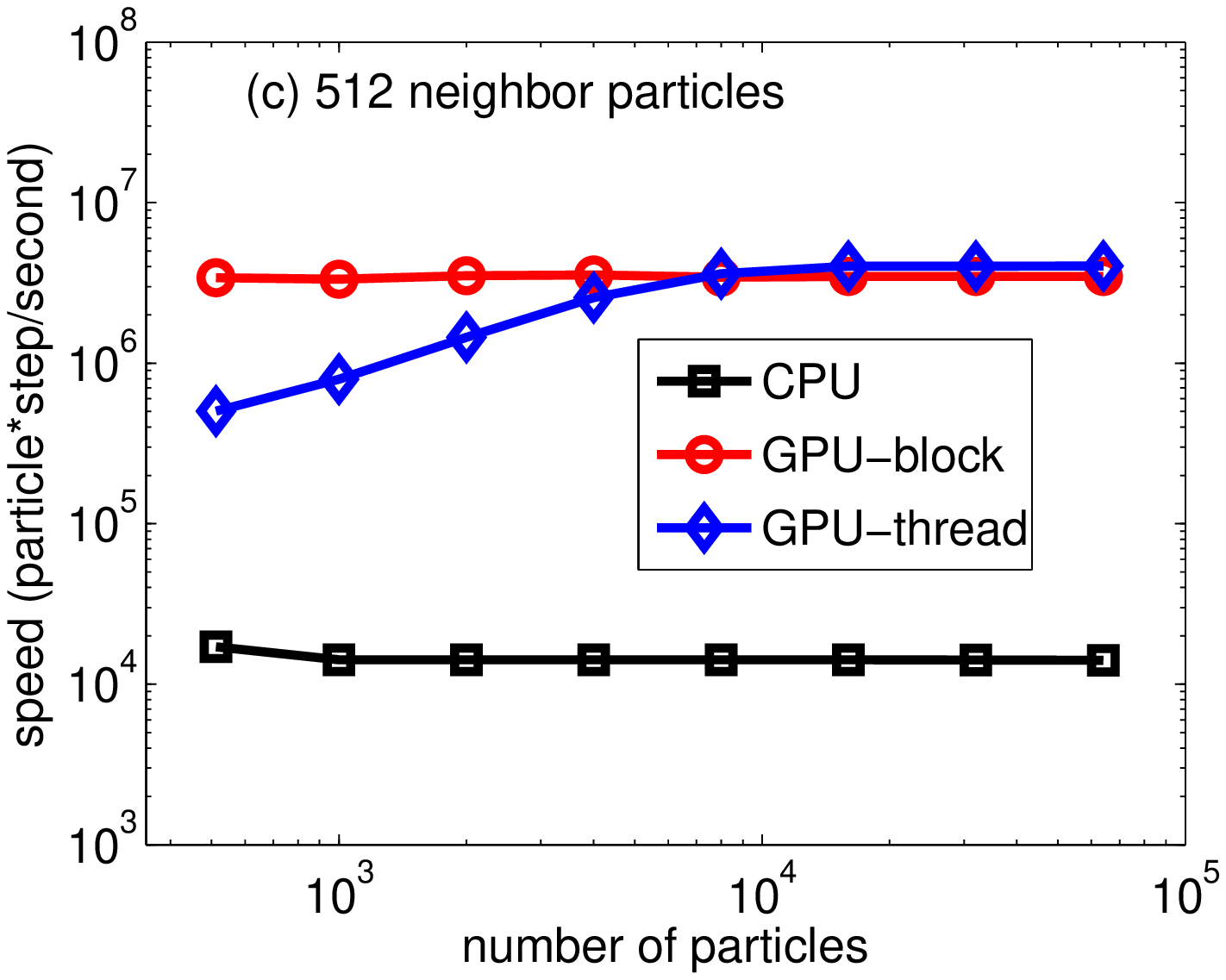}\\
  \includegraphics[width=2.2 in]{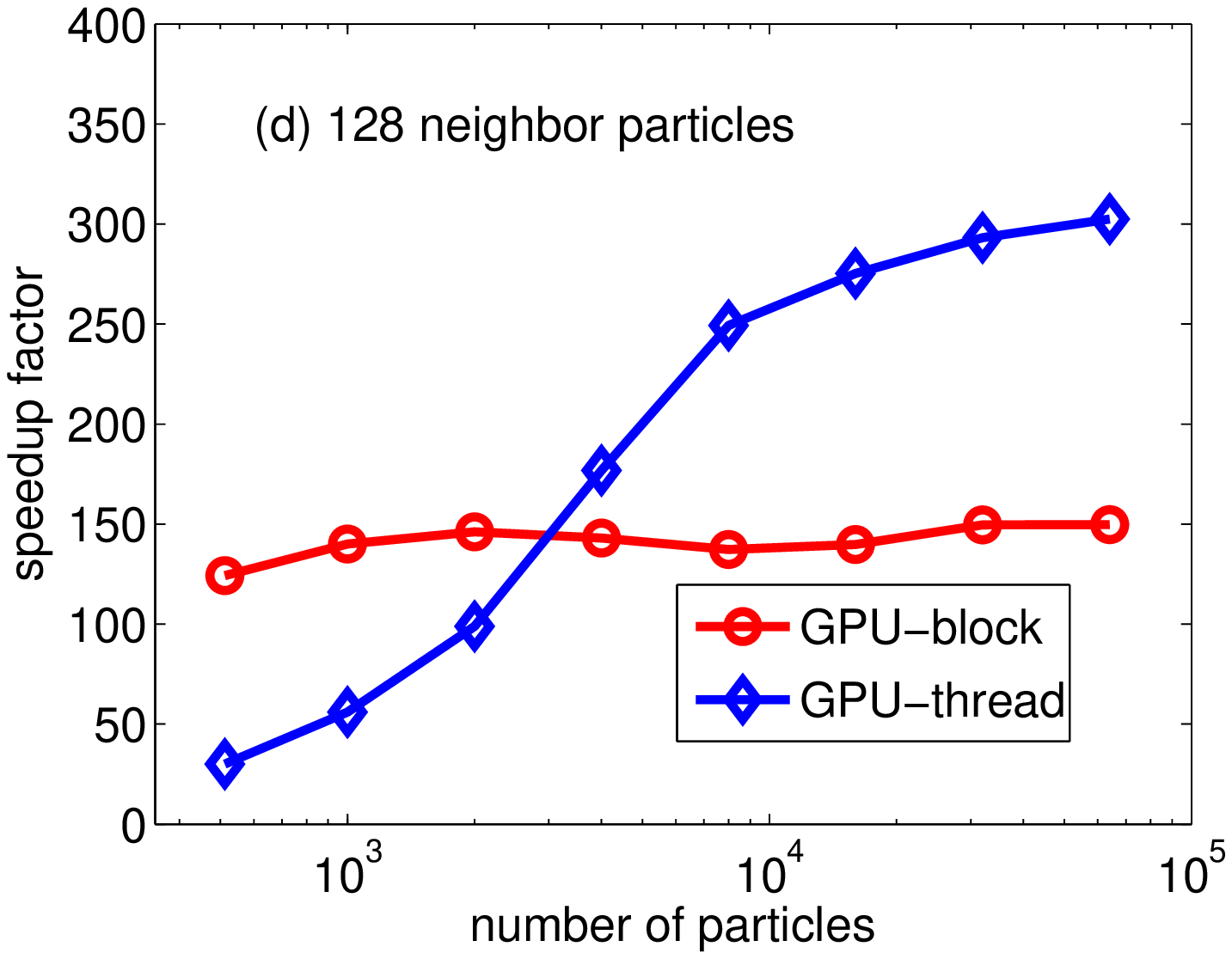}&
  \includegraphics[width=2.2 in]{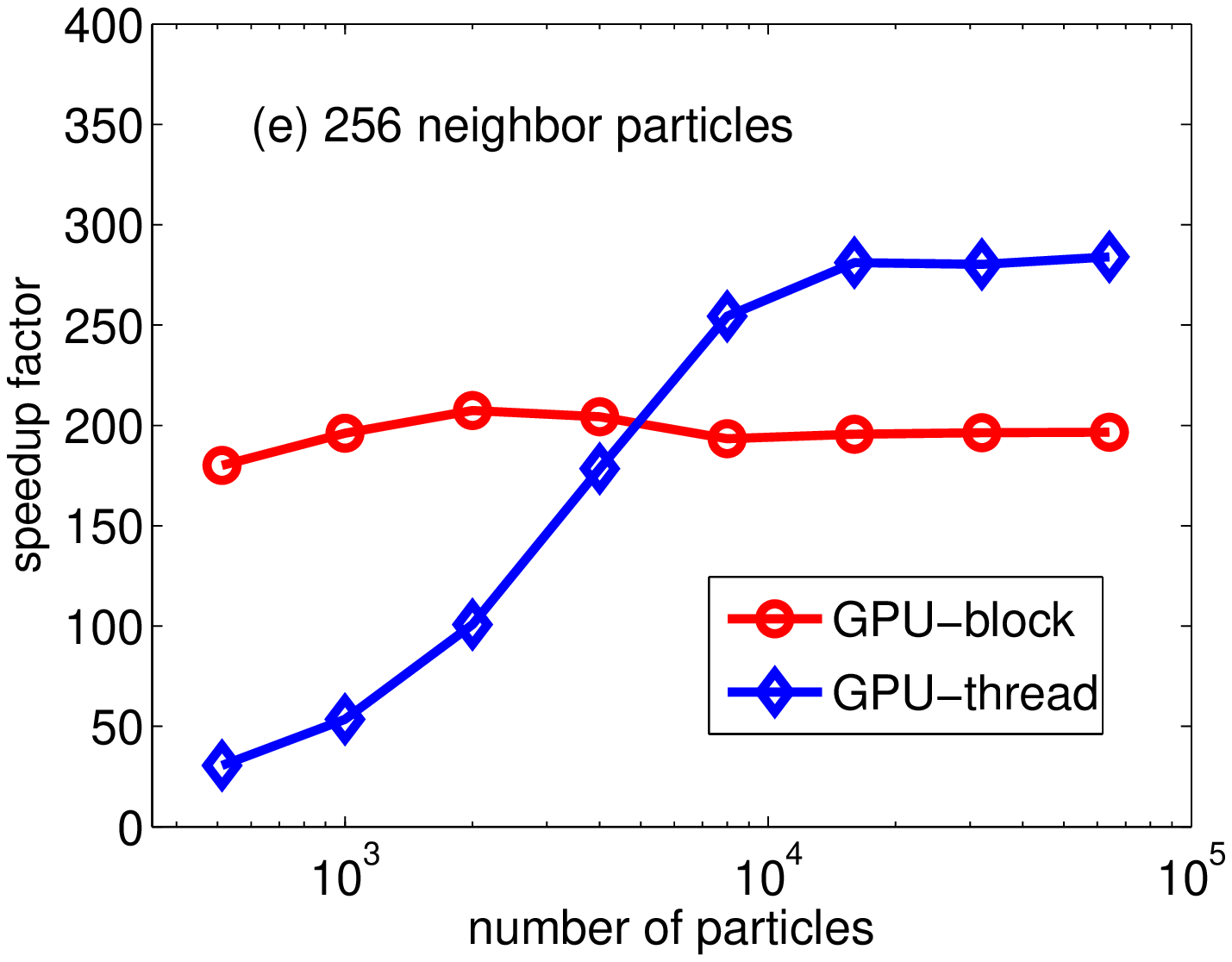}&
  \includegraphics[width=2.2 in]{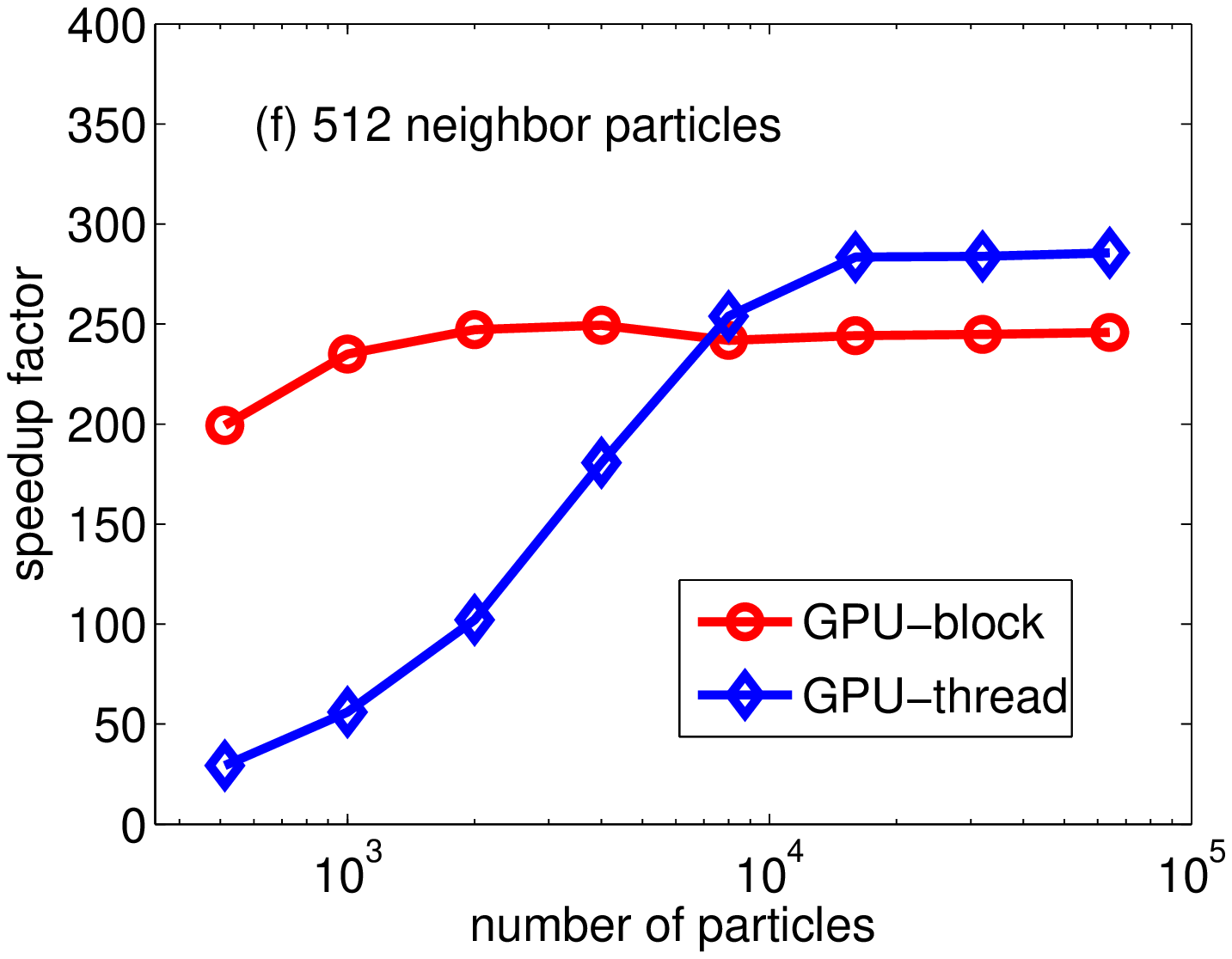}\\
  \includegraphics[width=2.2 in]{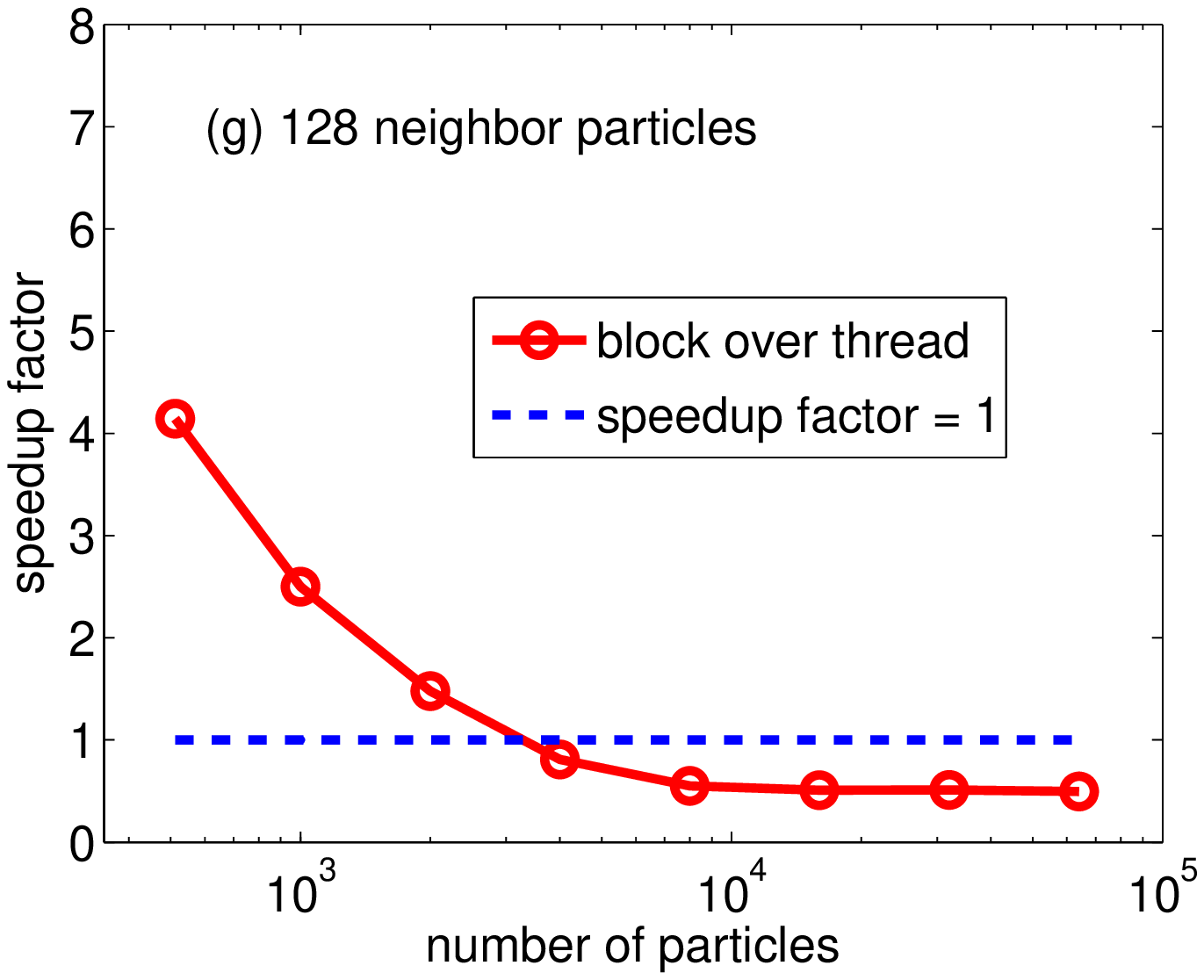}&
  \includegraphics[width=2.2 in]{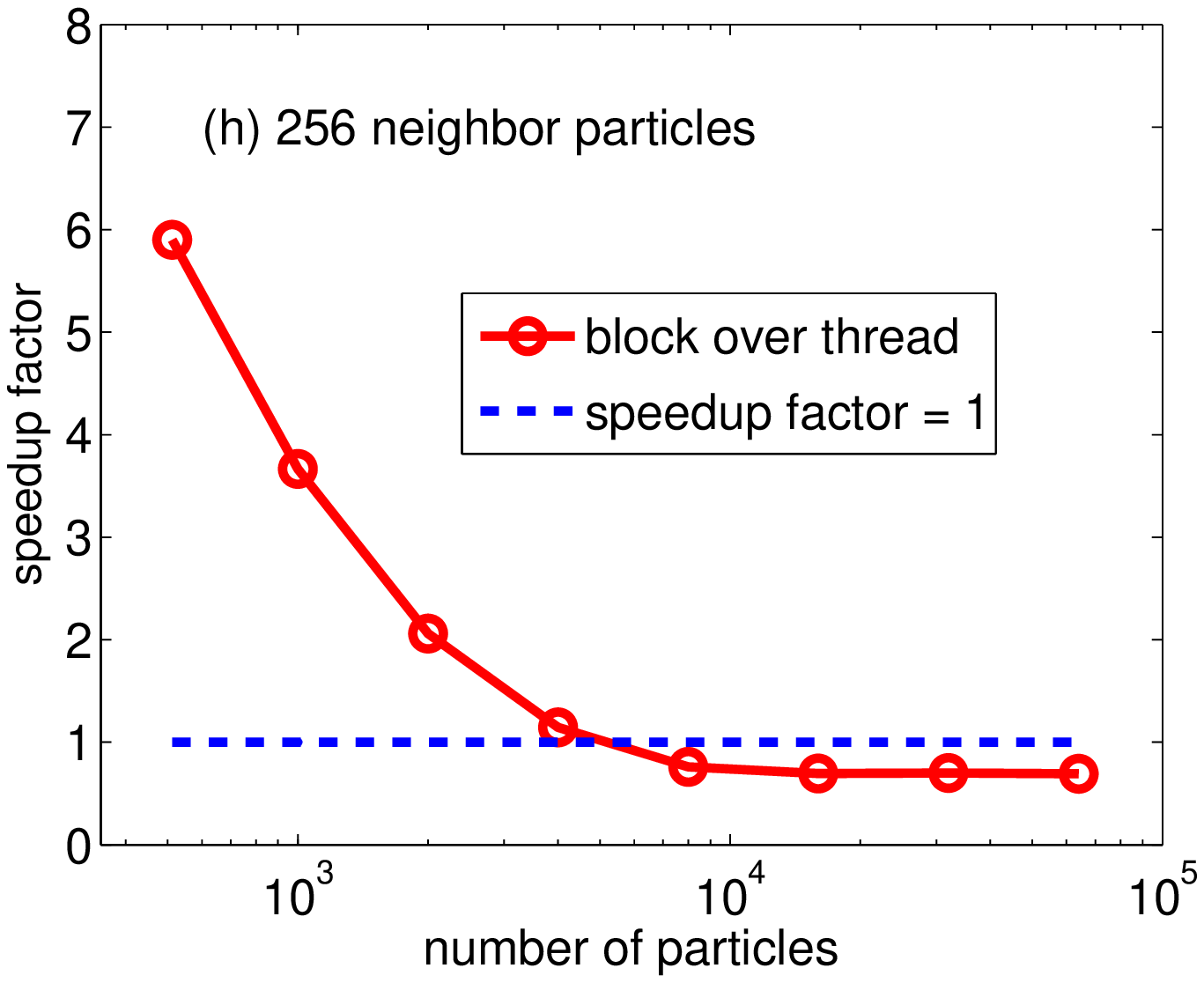}&
  \includegraphics[width=2.2 in]{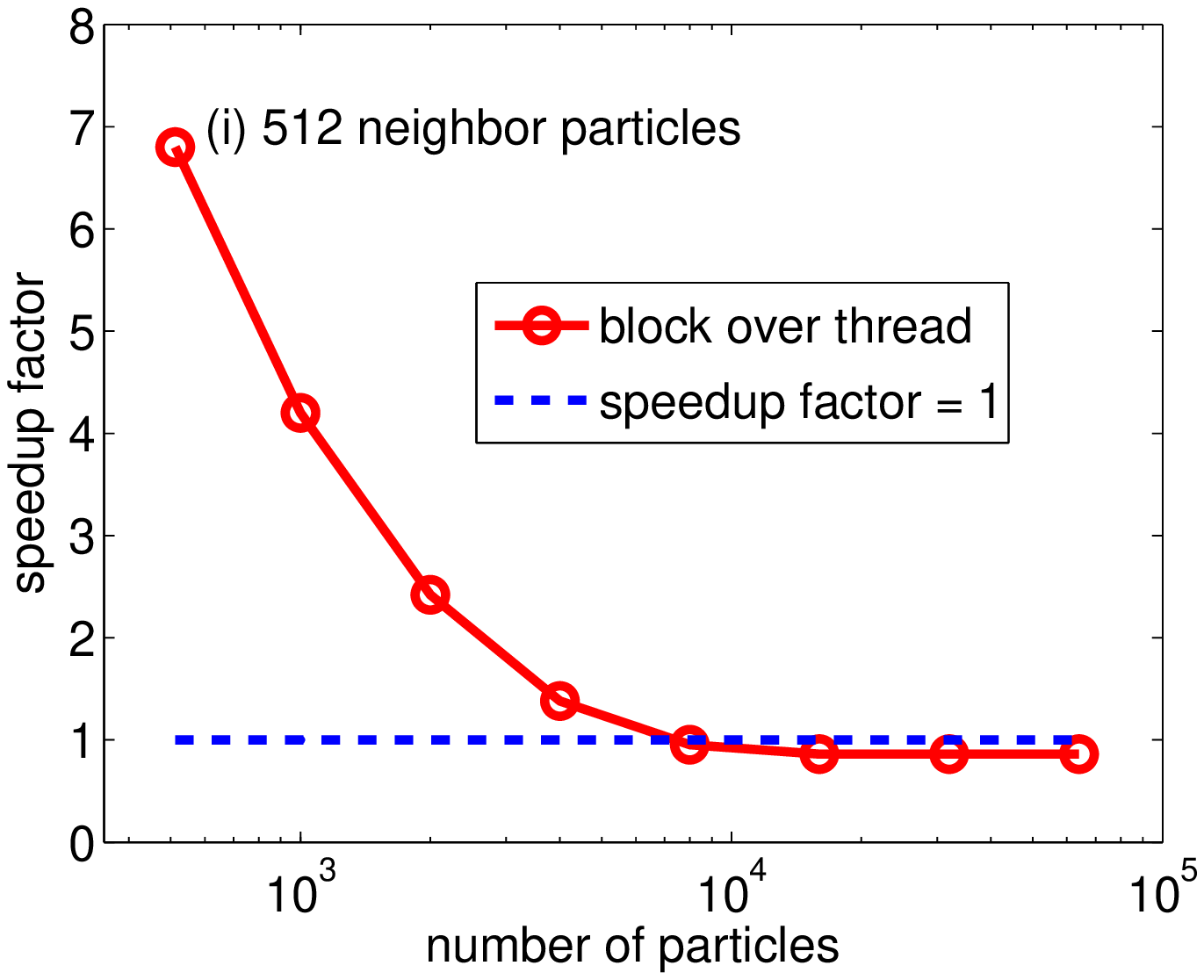}\\
\end{tabular}
  \caption{(Color online) Performance evaluation for the evolution part in the production stage
           for the RI potential: (a-c) computation speeds for both the CPU code 
           and the GPU code with different force evaluation schemes; (d-f) speedup
           factors of the GPU code over the CPU code; (g-i) relative speedup factors
           of the block-scheme over the thread-scheme. 
           The maximal number of neighbor particles
           in the simulations (128, 256 or 512) 
           are indicated in each subplot.
}
\label{figure:pbte_performance}
\end{center}
\end{figure*}

In this subsection, we evaluate the performance of the evolution part,
with an emphasis on the comparison of the two force evaluation
schemes.  To be specific, we only present the results for
the production stage, where the heat current data need to be calculated
and recorded. The results for the equilibration stage are similar.
We test the performance for the evolution part by measuring the computation
time for 1000 time steps. The computational speed is defined as the product
of the number of particles and the number of steps divided
by the computation time. In the literature \cite{anderson2008,rapaport2011},
the inverse of this quantity is also used to evaluate the computational
speed. The speedup factor is defined to be the computation time for the CPU
over that for the GPU. The system size used for our evaluation spans from
a few hundred to 64 thousand particles. We also considered different cutoff radii,
which determine the maximal number of neighbor particles $N_m$.

\subsubsection{The LJ potential}

The performances for the LJ potential are presented in
Fig. \ref{figure:argon_performance}, where the computational speeds
and the speedup factors are plotted against the simulation size.

For our CPU code, the LJ potential with $N_m = 128$
executes at a speed of about $5.0\times 10^5$ \text{particle} $\cdot$ step / second,
or equivalently, 2.0 $\mu$s / (particle $\cdot$ step).
The computational speed decreases with the increasing of
the maximal number of neighbor particles $N_m$. For
$N_m = 512$, the computational speed slows down to about
$9.0\times 10^4$ \text{particle} $\cdot$ step / second.
For relatively small systems ($N < 10^4$), the computation speed is nearly
independent of the simulation size, indicating a good linear-scaling
dependence of the computation time on the simulation size.
However, the linear-scaling behavior is not well preserved for
relatively large systems ($N > 10^4$). This is probably due to
the more expensive memory operations for larger data arrays
associated with larger simulation size. We will come to this
problem when we discuss the RI potential later.

Our GPU implementation achieves high performance and
large speedup factors. By using the thread-scheme, the
computational speed can be as high as
$3.5\times 10^7$ \text{particle} $\cdot$ step / second,
or equivalently, 0.029 $\mu$s / (particle $\cdot$ step)
for $N_m = 128$ and $N = 8000$, which gives a speedup
factor of about 70.  For $N_m = 512$ and $N = 8000$, the speedup factor
can reach 95 by using the thread-scheme.

From Fig. \ref{figure:argon_performance} (a-c)
we can see that the computational speeds for the
thread-scheme and the block-scheme saturate
at different simulation sizes. As discussed before,
the  number of invoked blocks for the force evaluation kernel
in the thread-scheme is  $\lceil N/S_b \rceil$. With a
block size of $S_b=128$, the number of blocks only reaches
the upper bound of the number of resident blocks in the GPU
when $N=16384$. This explains why the performance for the
thread-scheme saturates at about $N=10^4$.
In contrast, the number of invoked blocks for the force
evaluation kernel in the block-scheme is the number of particles,
and the block-scheme can attain its peak performance with only
a few hundred particles.

There is always a crossover of the performances for
the two force evaluation schemes. For the cases of $N_m$ = 128,
256 and 512, the simulation sizes at which the crossovers take
place are around 1000, 3000 and 5000, respectively.
From Fig. \ref{figure:argon_performance} (g-i) we can see that 
while the thread-scheme is a little faster when $N>10^4$,
the block-scheme is several times faster when $N<10^4$. 
The smaller the system, the higher the relative speedup factor
of the block-scheme over the thread-scheme.
This makes the block-scheme
more preferable for thermal conductivity calculation using the
Green-Kubo method, where we rarely need to
consider system with more than a few thousand particles.
There are two main reasons for the  superior performance
of the thread-scheme over the block-scheme
for large systems when saturation is obtained for both schemes.
The first is that in the block-scheme,
the total force and heat current for each particle has to be
accumulated by binary reduction, during which only a portion of threads
are used to do the calculation. The second is that there is more global memory
access for the block-scheme. In the block-scheme, the positions
$\mathbf{r}_i$ and velocities $\mathbf{v}_i$ as used in lines 5 and 6 of
\textbf{Algorithm \ref{algorithm:force_block}} need to be transfered
from global memory to all the threads in the block corresponding to
the heading particle $i$. In contrast, in the thread-scheme, the positions
and velocities for a heading particle need only to be transfered from
global memory to a single thread.

From Fig. \ref{figure:argon_performance} (d-f) we can see that
while the speedup factors for the thread-scheme nearly do not vary
with the increasing of $N_m$, the speedup factors for the block-scheme
increase significantly with the increasing of $N_m$. The
computation time for the thread-scheme is dominated by the \textbf{for} loop
in lines 4-12 of \textbf{Algorithm \ref{algorithm:force_thread}} and
scales nearly linearly with respect to $N_m$, as in the case of the CPU code.
In contrast, in the block-scheme, a significant portion of computation time is
spent on the extra global memory access of positions and velocities and
the binary reduction operations for force and heat current.
The amount of the extra global memory access scales with $N$ and
the number of binary reduction operations is determined by $S_b$,
both of which do not scale with $N_m$. Thus, the block-scheme performs
better and better with the increasing of $N_m$.

Lastly, we note that there is an abrupt `drop' of the computational speed
for the thread-scheme around $N = 10^4$. This is likely caused
by increasingly inefficient use of the cache memories and the saturation
of the GPU's cores.

\subsubsection{The RI potential}

The RI potential is much more computationally intensive than the LJ potential.
One can see from Fig. \ref{figure:pbte_performance} (a) that for $N_m$ = 128,
the CPU computational speed for the RI potential is about
$6\times 10^4$ \text{particle} $\cdot$ step / second,
or equivalently, 17 $\mu$s / (particle $\cdot$ step), which is about 12 times slower
than that for the LJ potential with the same $N_m$.
Thus, compared to the LJ potential, the computation time for
the RI potential would be dominated by the actual floating point operations,
and this can help to understand why the linear-scaling behavior with respect
to the simulation size is preserved much better for the RI potential
(Fig. \ref{figure:pbte_performance} (a-c)) than
for the LJ potential (Fig. \ref{figure:argon_performance} (a-c)).

As for the performances for the GPU code, the results for the RI potential
are even more impressive compared with the LJ potential.
The speedup factors can be as large as
300 for the thread-scheme. The higher acceleration rate for the
RI potential compared to that for the LJ potential results
from the higher arithmetic intensity of RI potential.

The testing results for the RI potential exhibit similar features as in
the case of the LJ potential, which can be listed as follows:
\begin{enumerate}
 \item The performance for the thread-scheme saturates only when $N>10^4$,
       while that for the block-scheme saturates for a few hundred particles.
 \item For each $N_m$, there is a crossover point for the performance curves,
       before which the block-scheme performs better, and after which the opposite
       is true. The simulation sizes associated with the crossover points
       are around 2000, 5000 and 7000 for $N_m$ = 128, 256 and 512, respectively.
 \item The performance for the thread-scheme is weakly dependent on the value
       of $N_m$. In contrast, the average speedup factor for the block-scheme
       varies from 150 to 200 to 250 when $N_m$ increases from 128 to 256 to 512.
\end{enumerate}
All these points can be explained similarly as in the case of the LJ potential.

\subsection{Neighbor list construction}

\begin{figure}
\begin{center}
  \includegraphics[width=3 in]{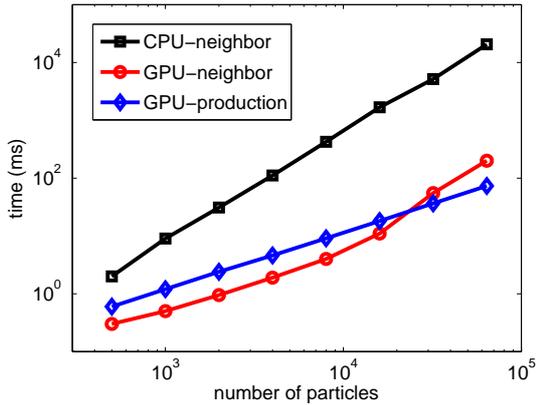}
  \caption{Computation times for the neighbor list construction in the CPU
           and the GPU as a function of the simulation size.
           The computation time of 10 production steps without
           neighbor list update in the block-scheme for LJ potential with
           $N_m$ = 128 is also presented for comparison.}
  \label{figure:neighbor}
\end{center}
\end{figure}

The computation times for constructing the neighbor
list in the CPU and the GPU are compared
in Fig. \ref{figure:neighbor}.
The computation time for the CPU implementation scales quadratically with the
simulation size, as expected. Although our GPU implementation is nearly
a direct translation of the CPU implementation, its computation time scales
linearly for $N < 10^4$ and only begin to scale quadratically for $N > 10^4$.
Furthermore, for $N < 10^4$, the computation time for the neighbor list construction
is less than that of 10 production steps  without neighbor list
update in the block-scheme for LJ potential with $N_m$ = 128.
As a comparison, the computation time for the neighbor list construction reported by
Anderson \textit{et al} \cite{anderson2008}, 
who use a cell decomposition approach, is about 10 times
larger than their reported computation time for the force evaluation with one step.
Thus, for $N < 10^4$, the simple neighbor list construction method as given in
\textbf{Algorithm \ref{algorithm:neighbor}} can lead to rather
good performance. However, for larger systems, the cell decomposition approach
will definitely perform better.

\subsection{HCACF calculation}

\begin{figure}
\begin{center}
  \includegraphics[width=3 in]{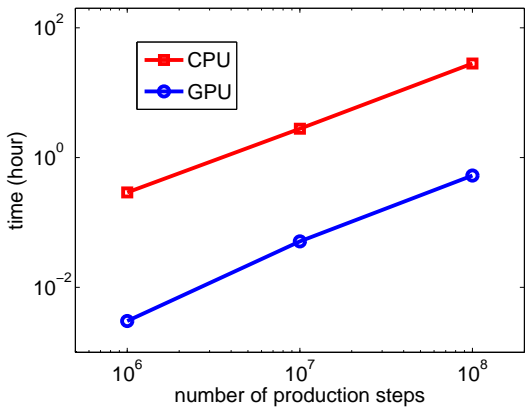}
  \caption{(Color online) Computation times for the HCACF calculations in CPU
           and GPU as functions of the production time steps $N_p$.
           The number of data points for HCACF is chosen to be
           $N_c=5\times10^4$.}
  \label{figure:hcacf}
\end{center}
\end{figure}

Figure \ref{figure:hcacf} presents the computation times for the HCACF
calculation in the CPU and the GPU. The calculation of HCACF for the case of
$N_c = 5 \times 10^4$ and $N_p = 10^8$ takes up more than one day using a CPU
but only half an hour using a GPU. From the previous discussion, we can
see that the evolution part of the MD simulation achieves more than two
orders of magnitude speedup. If we do not implement this post-processing
part in the GPU as well, the speedup factor for the whole program will
be much lower than that for the evolution part alone. 
For example, in the 
next section, we will apply the GPU code in the block-scheme to 
calculate the thermal conductivity of PbTe at a given
temperature using the following values of the relevant parameters:
$N = 512$, $N_m = 512$ and $N_p = 10^8$. The computational
speed is  $3.4 \times 10^6$ \text{particle} $\cdot$ step / second by using 
the block-scheme (Fig. \ref{figure:pbte_performance} (c)), and 
the computation time for the evolution part would be less than 5 hours, 
which is even much less than that for the HCACF calculation in CPU.

\section{Validation}
\label{section:Validation}

In this section, we validate our GPU implementation by studying the lattice
thermal conductivity of solid argon and bulk PbTe, using the LJ potential
and the RI potential, respectively.

\subsection{Determining the lattice constant at zero pressure}

\begin{figure*}
\begin{center}
\begin{tabular}{cc}
  \includegraphics[width=3 in]{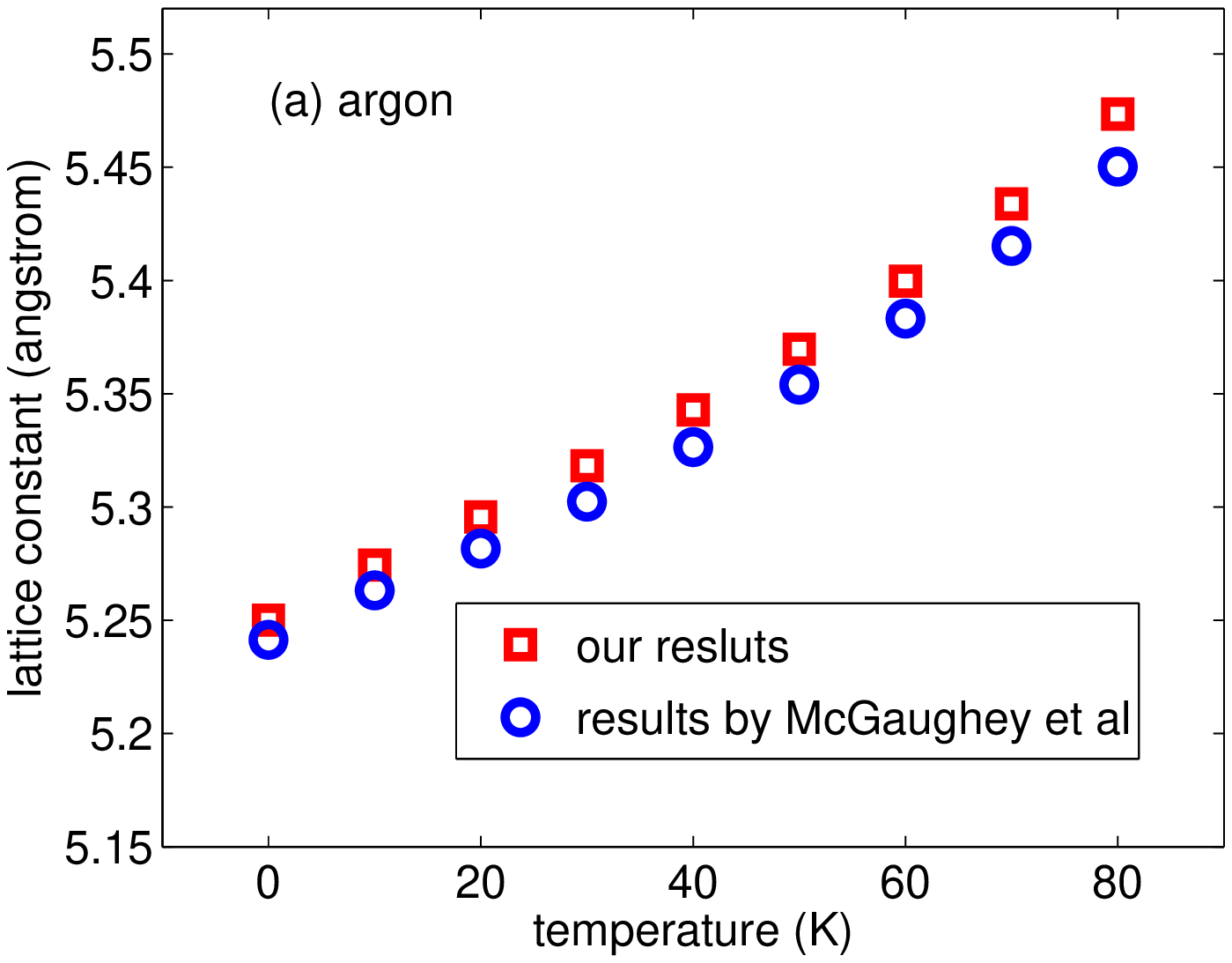}&
  \includegraphics[width=3 in]{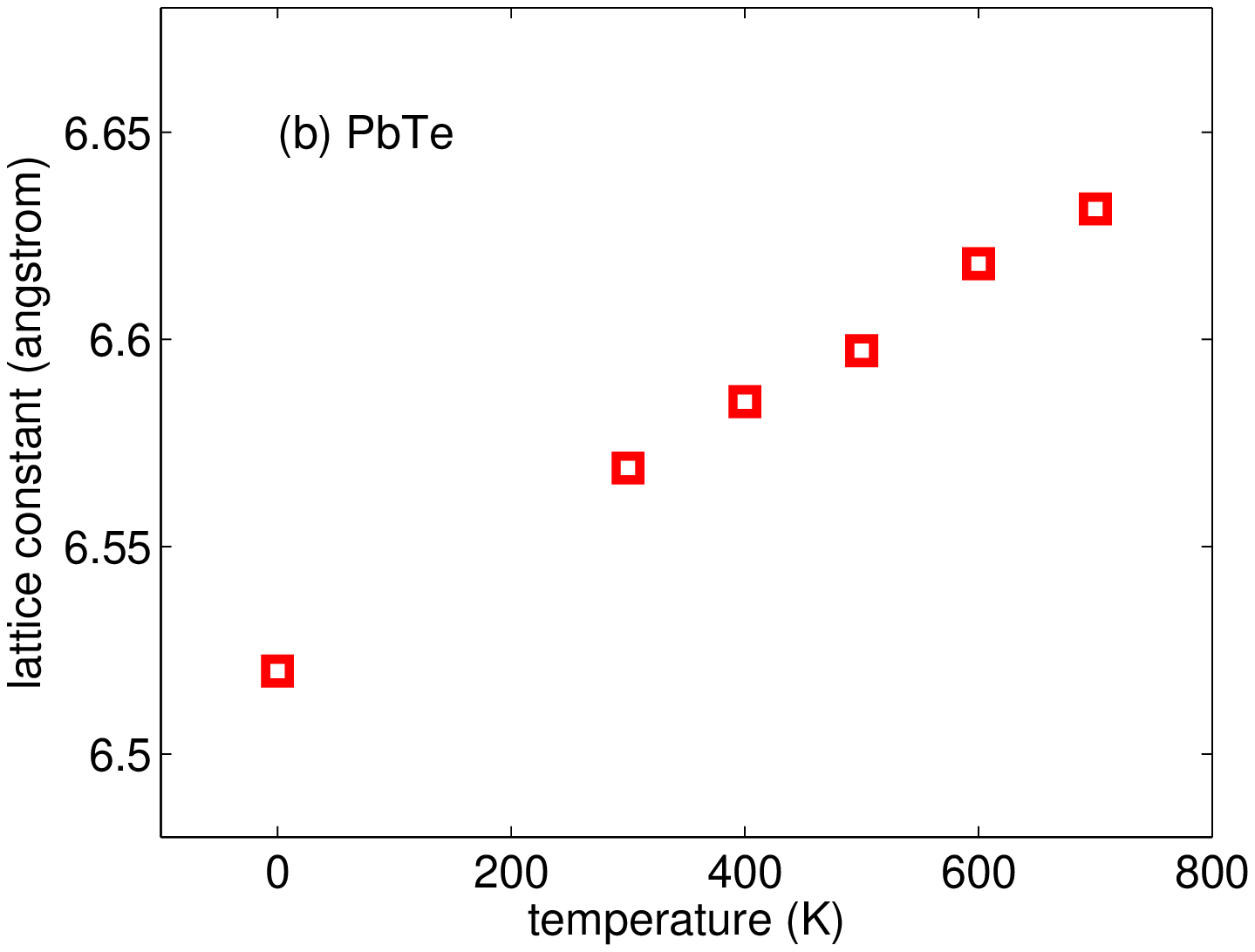}\\
\end{tabular}
  \caption{(Color online)
           Zero-pressure lattice constants for solid argon (a) and PbTe (b)
           at different temperatures. For solid argon, the results by McGaughey
           \textit{et al} \cite{mcgaughey2004} are also presented for comparison.}
  \label{figure:lattice_constant}
\end{center}
\end{figure*}

We only consider systems without external pressure (0 Pa).
The correct determination of the lattice
constant is crucial for the correct prediction of the
lattice thermal conductivity. In fact, the under-prediction of the
lattice thermal conductivities of solid argon \cite{kaburaki1999}
compared to experimental data results from an over-prediction of
the lattice constants, which is corrected by later studies
\cite{mcgaughey2004,tretiakova2004,chen2004,kaburaki2007}.
The zero-pressure lattice constant for non-bonded potential
can be obtained by using an NPT ensemble to control the pressure as well as
the temperature of a sufficiently large system with a large
cutoff radius for force evaluation.  The calculated lattice constants
at different temperatures for solid argon and PbTe are shown in
Fig. \ref{figure:lattice_constant}. For solid argon, the results
by McGaughey \textit{et al} \cite{mcgaughey2004} are also
presented for comparison. The lattice constants at zero temperature are
obtained from the cohesive energy curves. For solid argon, the zero-temperature
lattice constant is calculated to be 5.25 \AA. As a comparison, the
one obtained by McGaughey \textit{et al}  \cite{mcgaughey2004} is 5.24 \AA,  
and the experimental value \cite{ashcroft1976} is 5.30 \AA. For PbTe,
the zero-temperature lattice constant is calculated to be 6.52 \AA, which is
the same as that obtained by Qiu \textit{et al} \cite{qiu2008}. The lattice constants
for PbTe at elevated temperatures also exhibit a linear-dependence behavior on
temperature in the range of 300-700 K, from which we can deduce a
well defined value of the thermal
expansion coefficient, $2.30 \times 10^{-5}$ K$^{-1}$, which is
comparable to the experimental value \cite{houston1968},
$2.04 \times 10^{-5}$ K$^{-1}$.

\subsection{Results for lattice thermal conductivities}

After determining the lattice constants, we can calculate the zero-pressure
lattice thermal conductivities at different temperatures. 
For argon, the cutoff radius
for force evaluation is chosen to be $R_c = 3 \sigma$ and the time step
is chosen to be $\delta t = 2$ fs. We firstly equilibrate the 
simulated system in the NVT ensemble for $N_e = 10^6$ steps and then 
evolve the system in the NVE ensemble for $N_p = 10^7$ steps. 
The NVT ensemble in the equilibration stage serves to control the temperature
of the simulated system. After the system attains an equilibrated temperature, 
the heat current data are calculated and recorded at each time step in the NVE
ensemble, which is the most natural ensemble to simulate an equilibrated
system (with fluctuations, of course).
For PbTe, the corresponding parameters are chosen to be $R_c = 16$ \AA,
$\delta t = 0.2$ fs, $N_e = 10^7$ and $N_p = 10^8$. Another parameter 
which is only relevant for PbTe, namely, the electrostatic damping factor,
is chosen to be $\alpha = 0.2$ \AA$^{-1}$, 
which is a reasonable choice as suggested 
by Fennell \textit{et al} \cite{fennell2006}.

\subsubsection{Solid argon}

\begin{figure*}
\begin{center}
\begin{tabular}{ccc}
  \includegraphics[width=2.2 in]{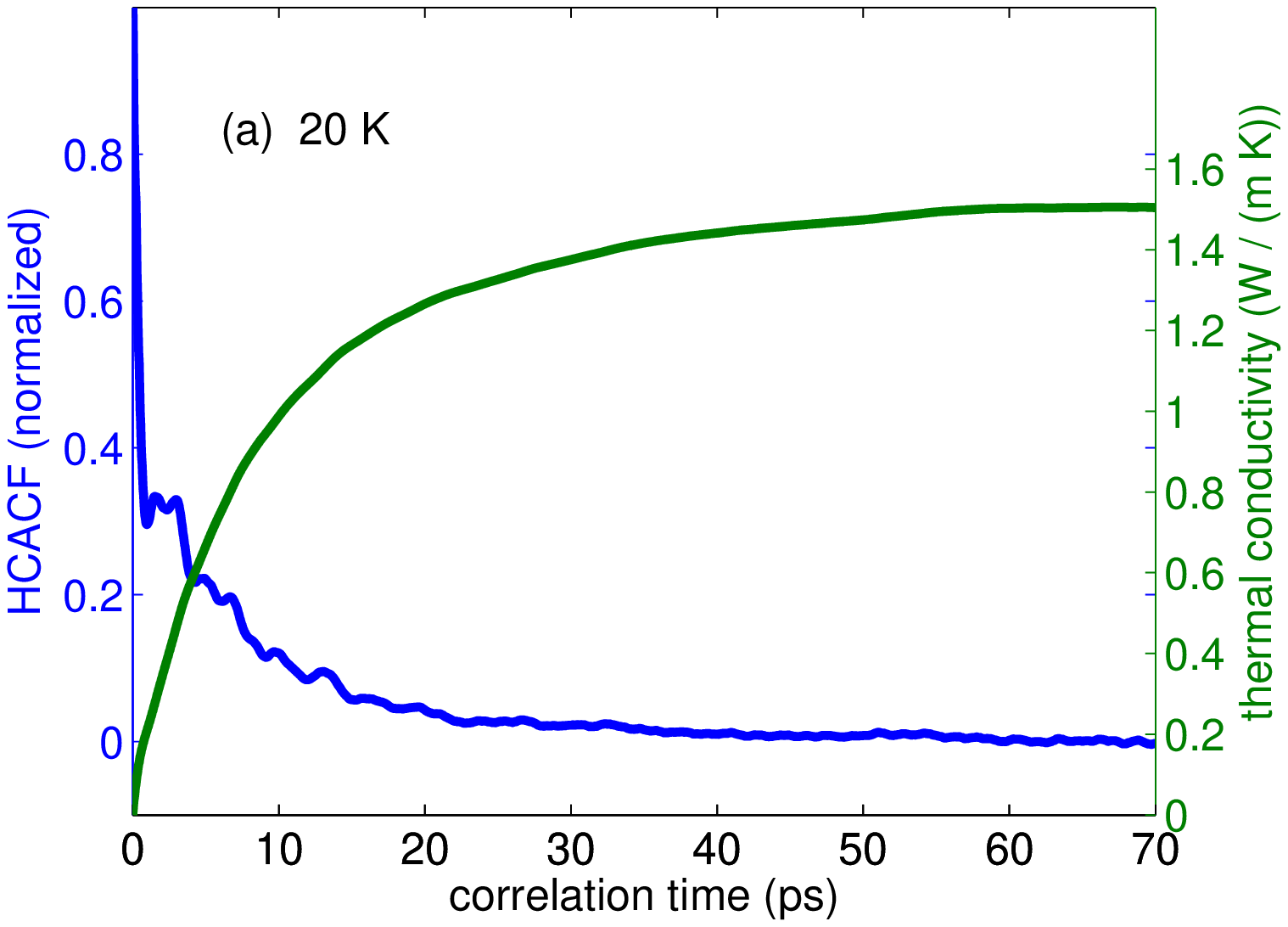}&
  \includegraphics[width=2.2 in]{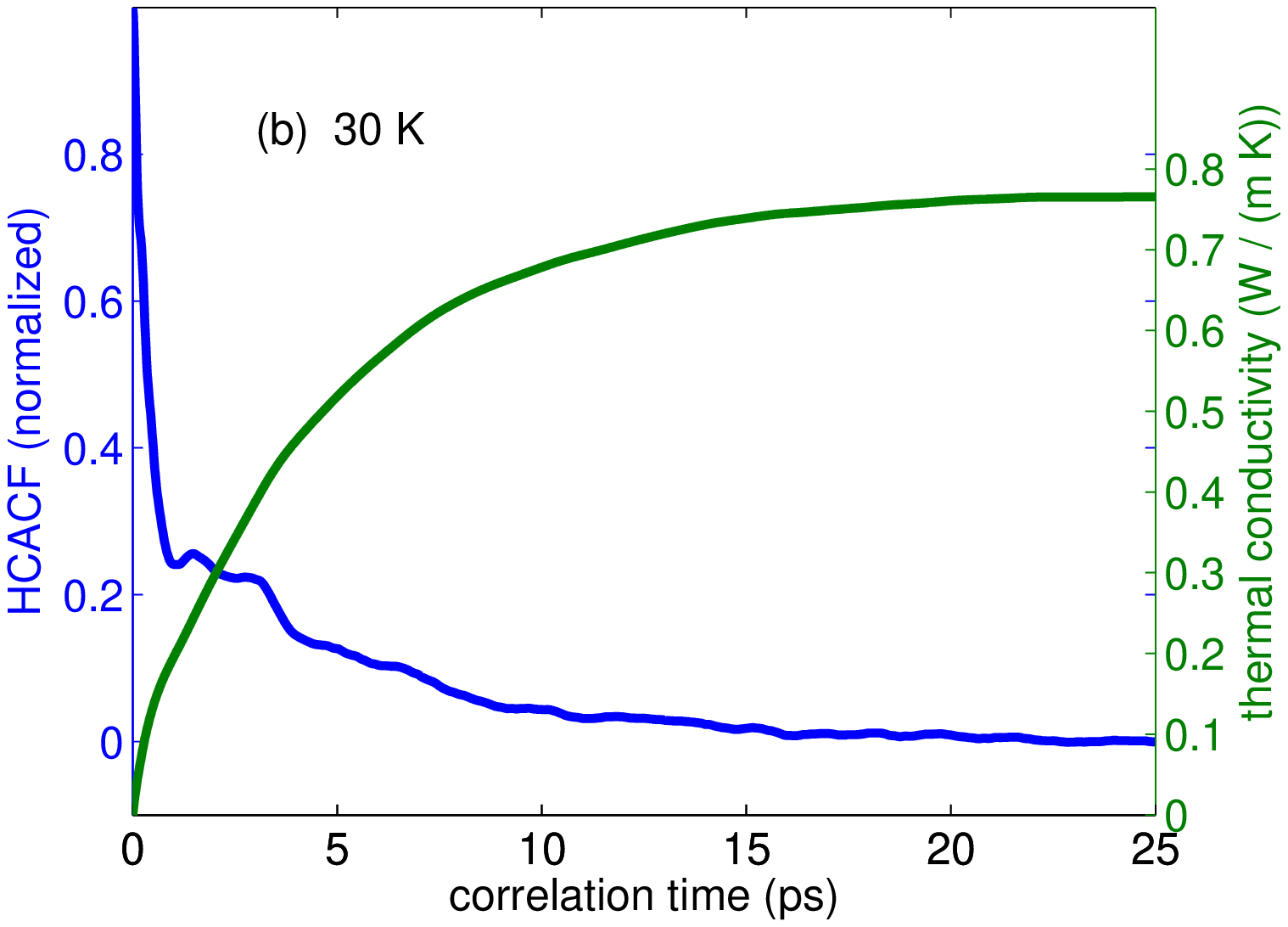}&
  \includegraphics[width=2.2 in]{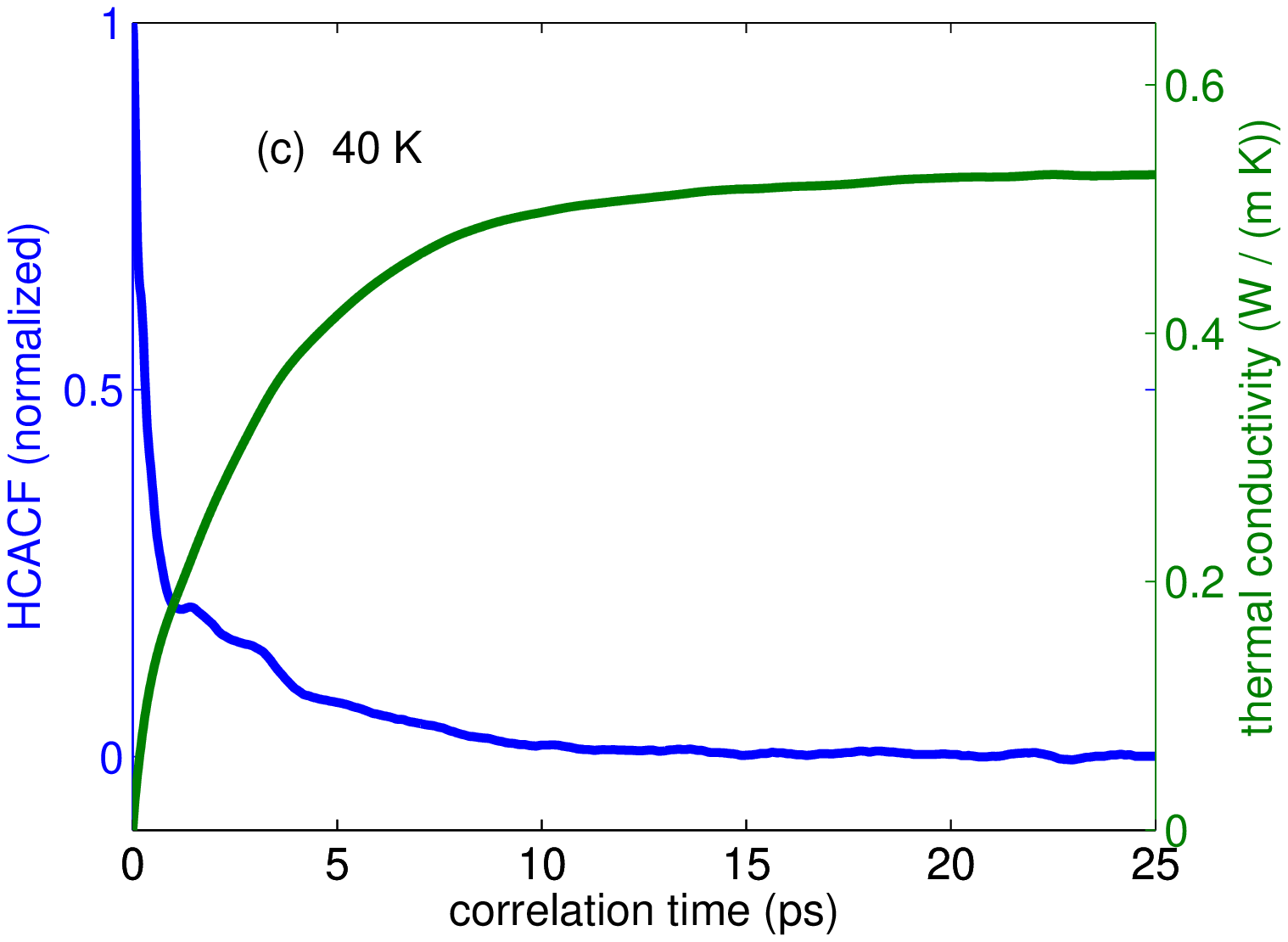}\\
  \includegraphics[width=2.2 in]{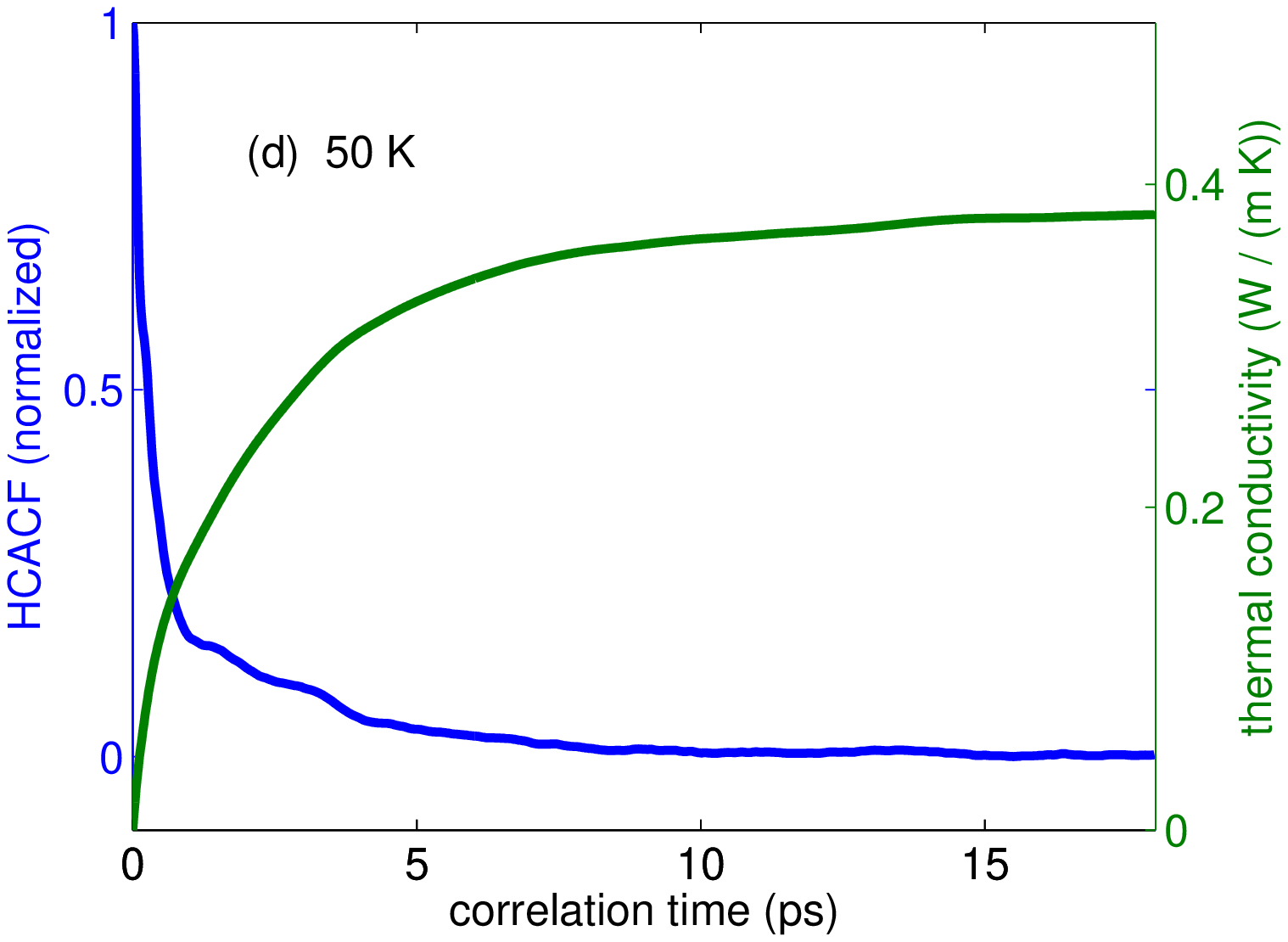}&
  \includegraphics[width=2.2 in]{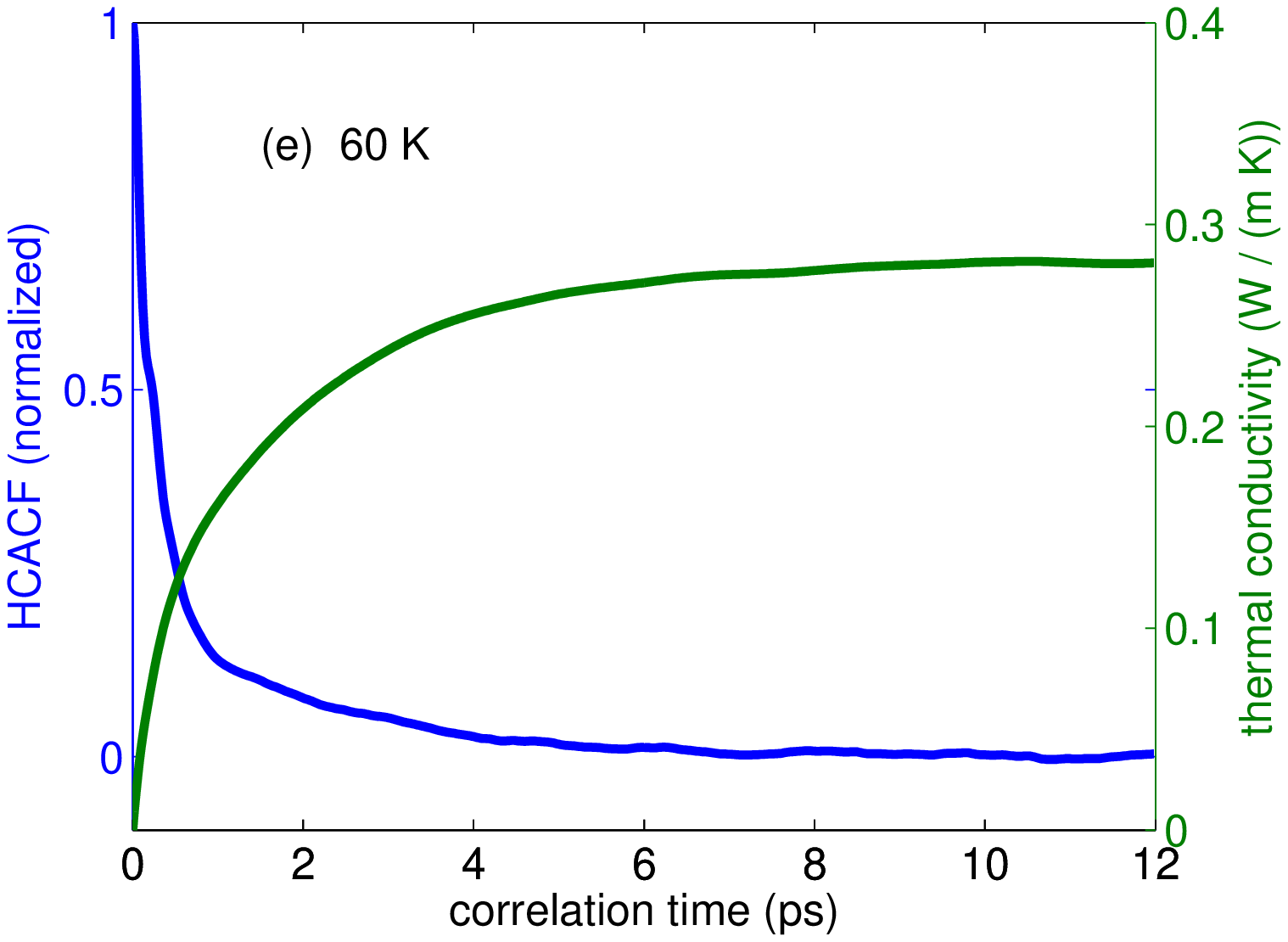}&
  \includegraphics[width=2.2 in]{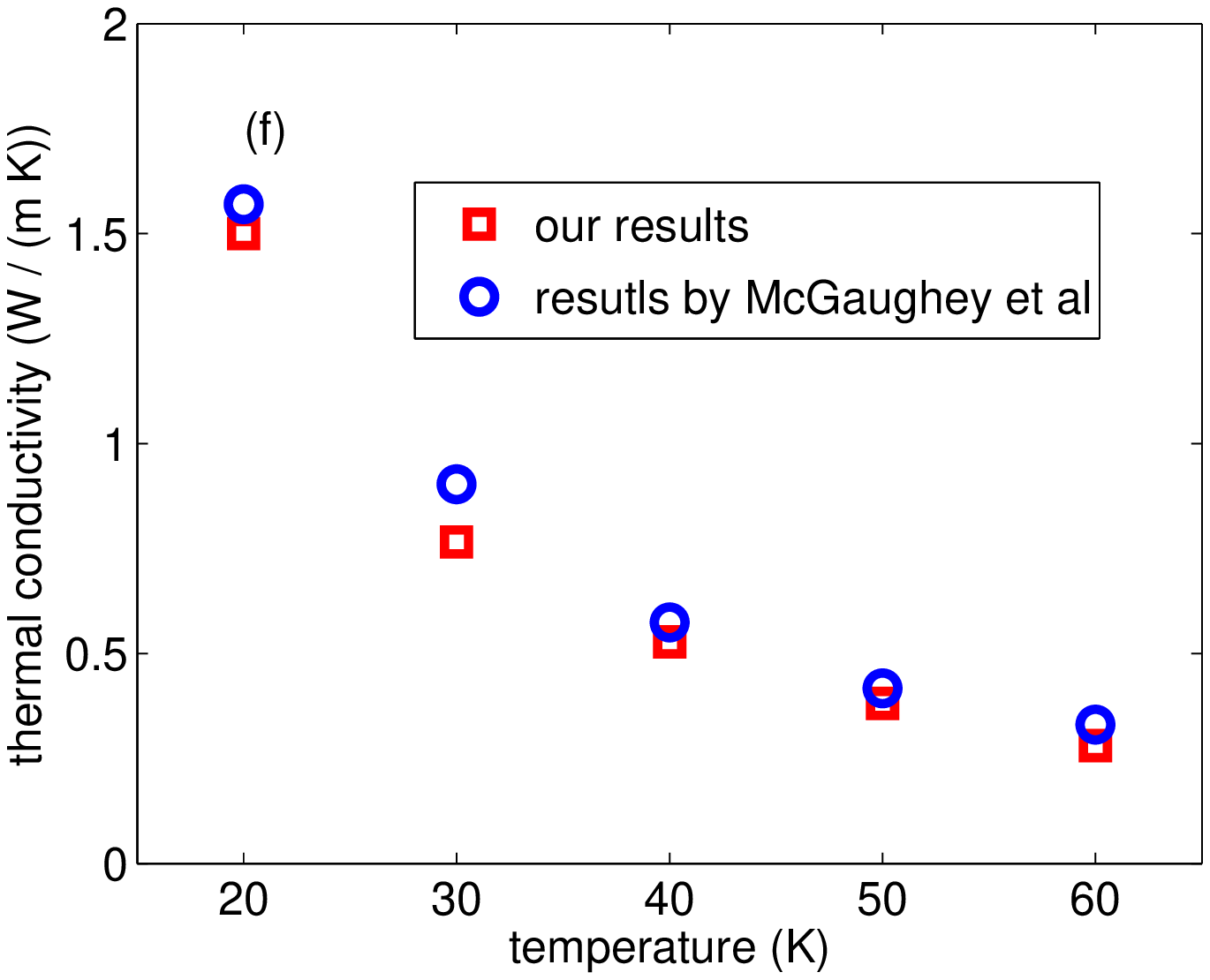}\\
\end{tabular}
  \caption{(Color online)
           HCACFs and RTCs (a-e) of solid argon as a function of the 
           correlation time and the converged lattice thermal conductivity as a function
           of temperature (f). In (f), the MD results obtained by
           McGaughey \textit{et al} \cite{mcgaughey2004} are also presented for comparison.}
  \label{figure:argon_kappa}
\end{center}
\end{figure*}

Figure \ref{figure:argon_kappa} (a-e) gives the calculated results for
HCACFs and RTCs at different temperatures
for solid argon using the GPU code in the block-scheme.
For all the temperatures, well converged HCACF and RTC
can always be obtained, as long as we collect sufficiently
many heat current data. The curves presented in Fig \ref{figure:argon_kappa} (a-e)
are obtained by setting the number of production steps to be  $N_p = 10^7$.
Since the curves are so smooth, we need not do any fitting to
obtain a well defined value of thermal conductivity.

Figure \ref{figure:argon_kappa} (f) compares our simulation results
with those reported by
McGaughey \textit{et al} \cite{mcgaughey2004}.
We have tested the finite size effects and found that
a simulation size of $N = 5\times5\times5\times4=500$ is sufficient.
It can be seen that our results agree well with theirs.
Since their results are well established,
this agreement provides a strong
evidence of the correctness of our program.

\subsubsection{PbTe}

\begin{figure*}
\begin{center}
\begin{tabular}{ccc}
  \includegraphics[width=2.2 in]{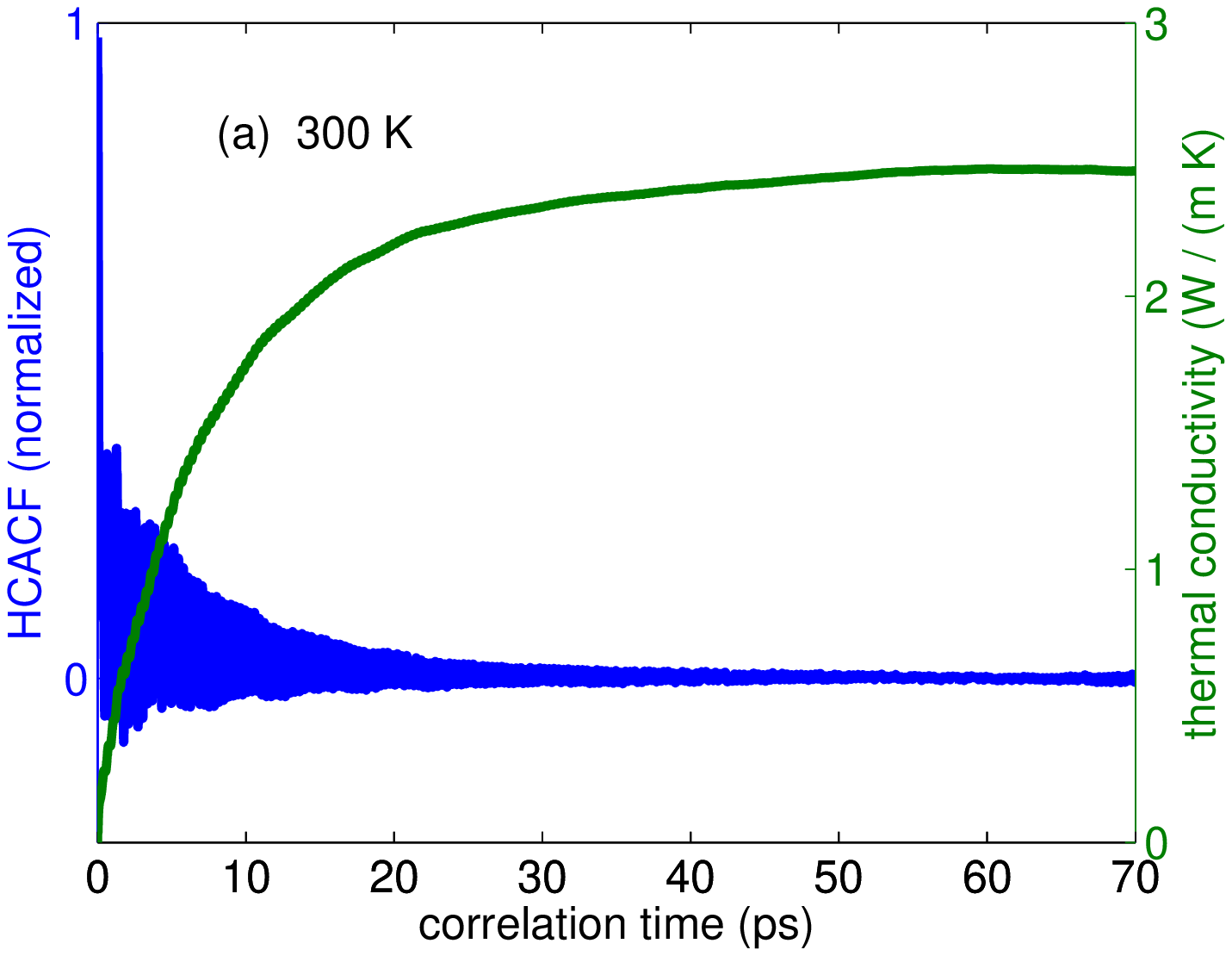}&
  \includegraphics[width=2.2 in]{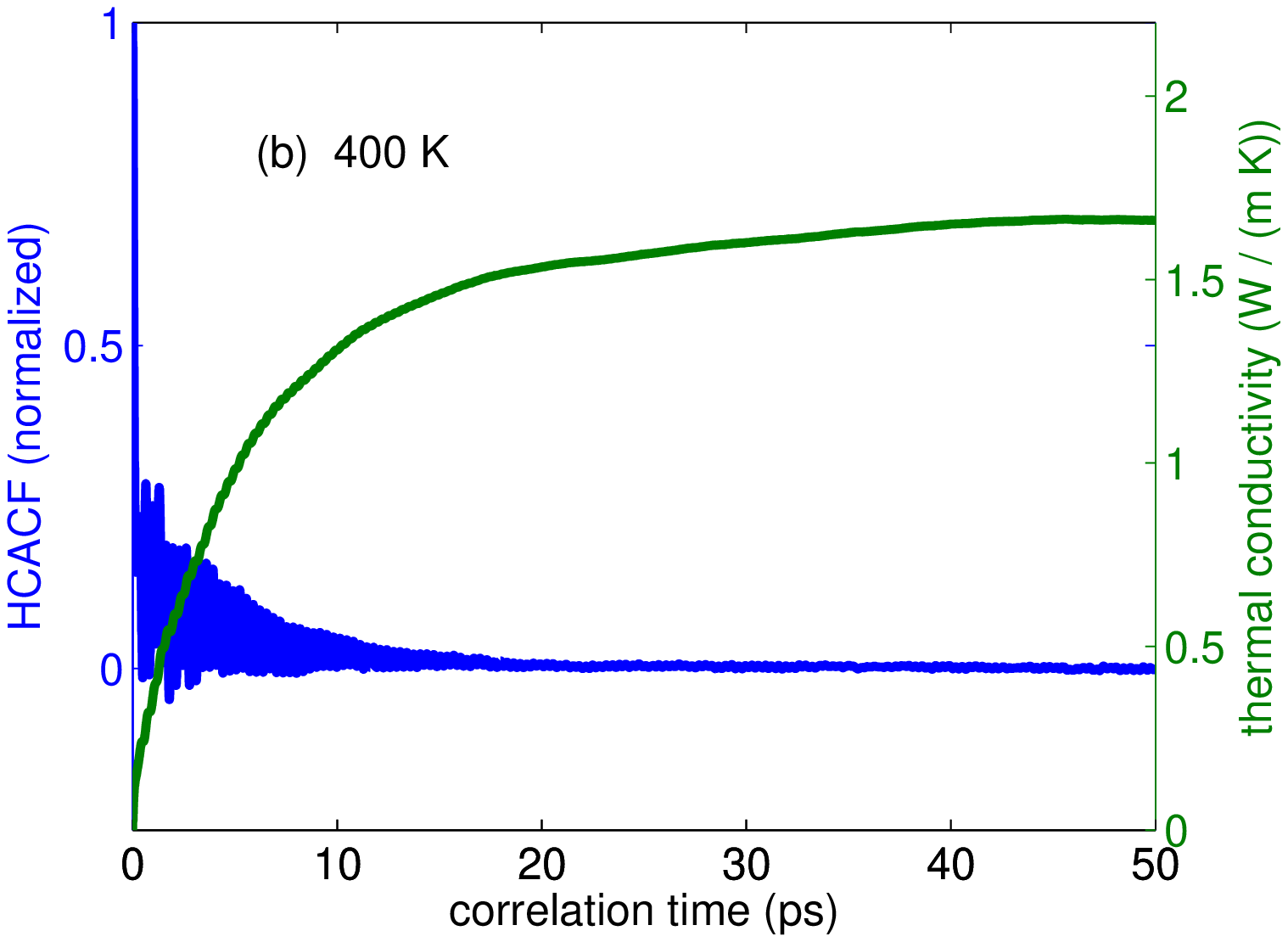}&
  \includegraphics[width=2.2 in]{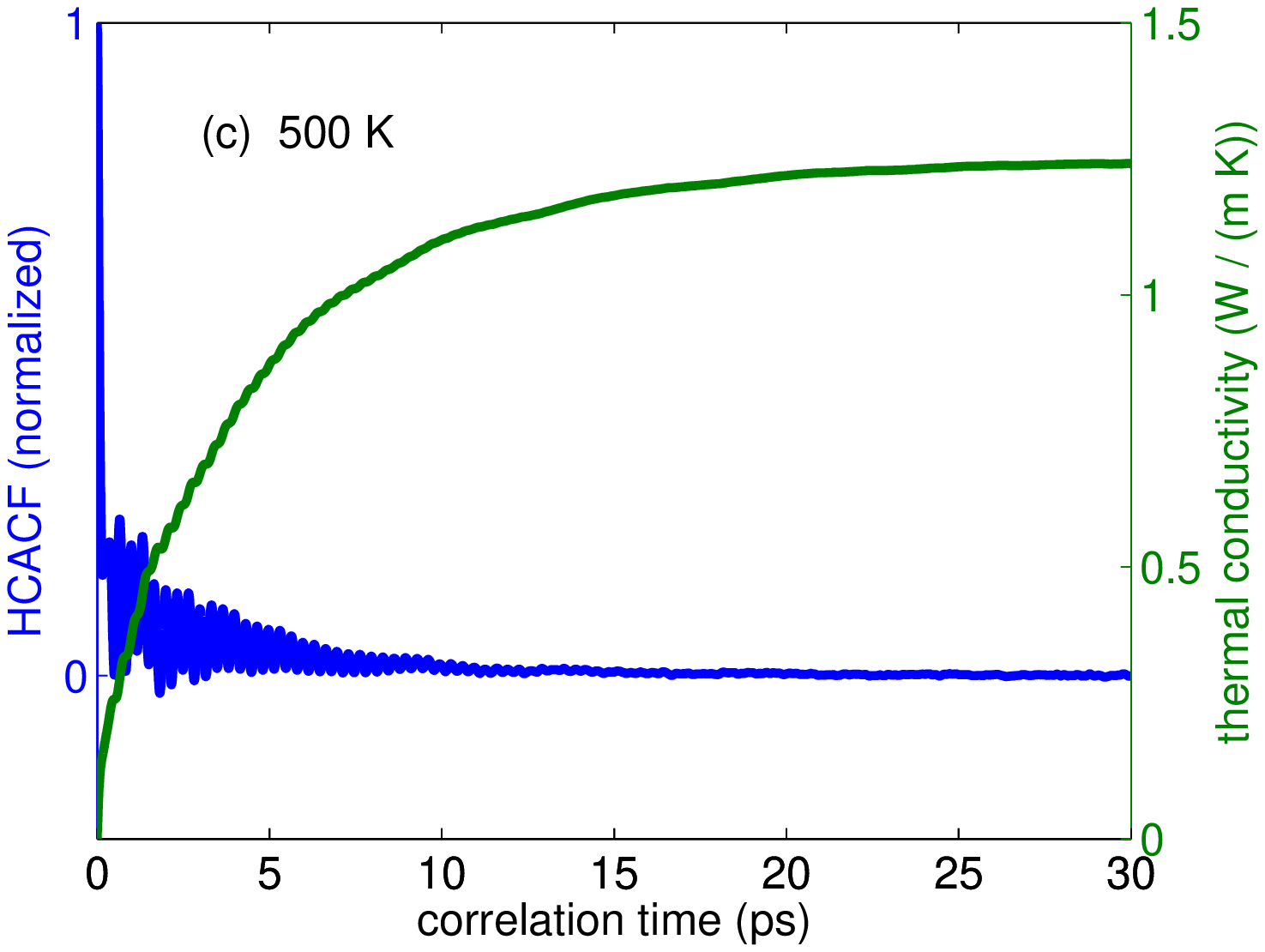}\\
  \includegraphics[width=2.2 in]{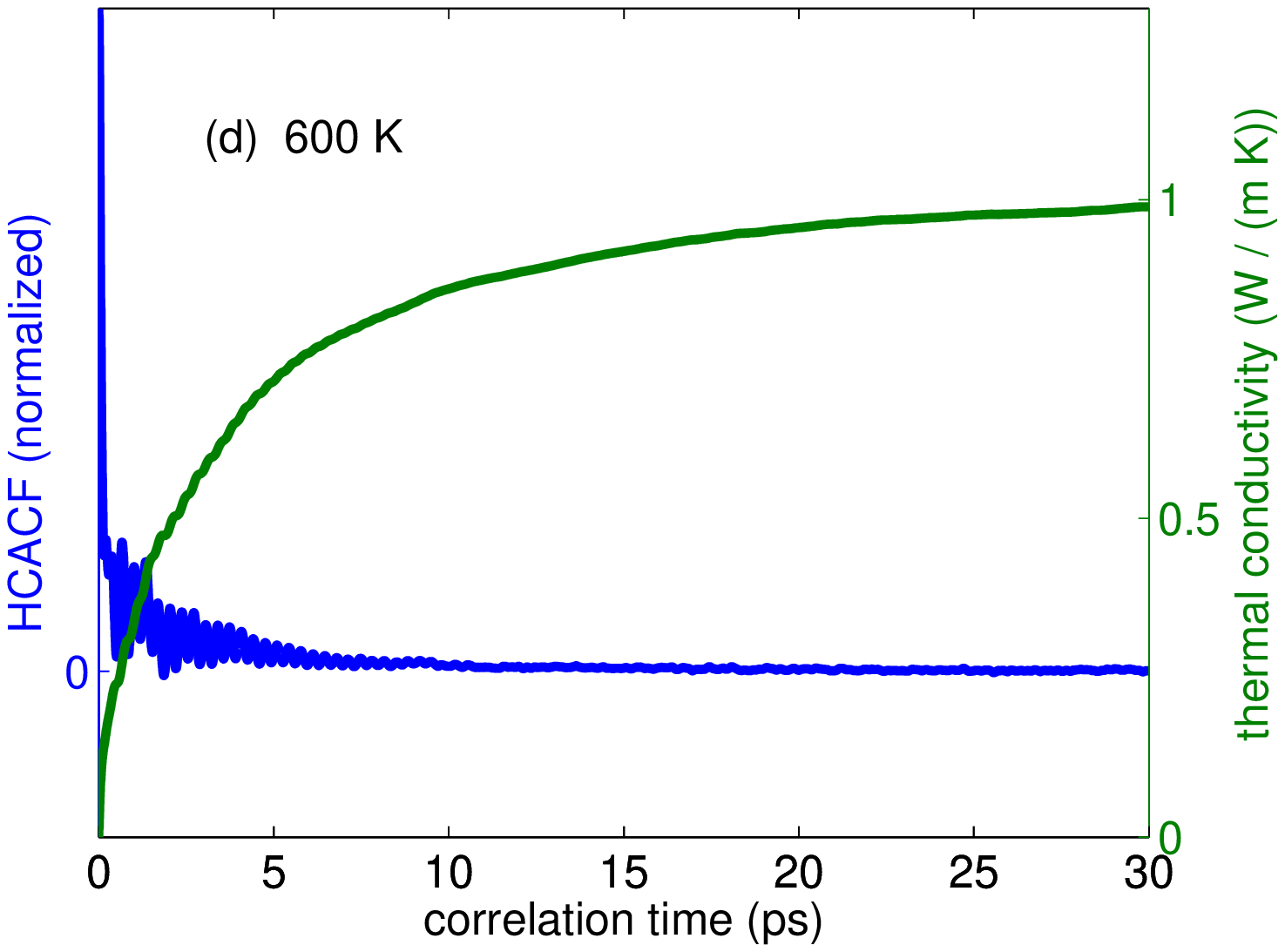}&
  \includegraphics[width=2.2 in]{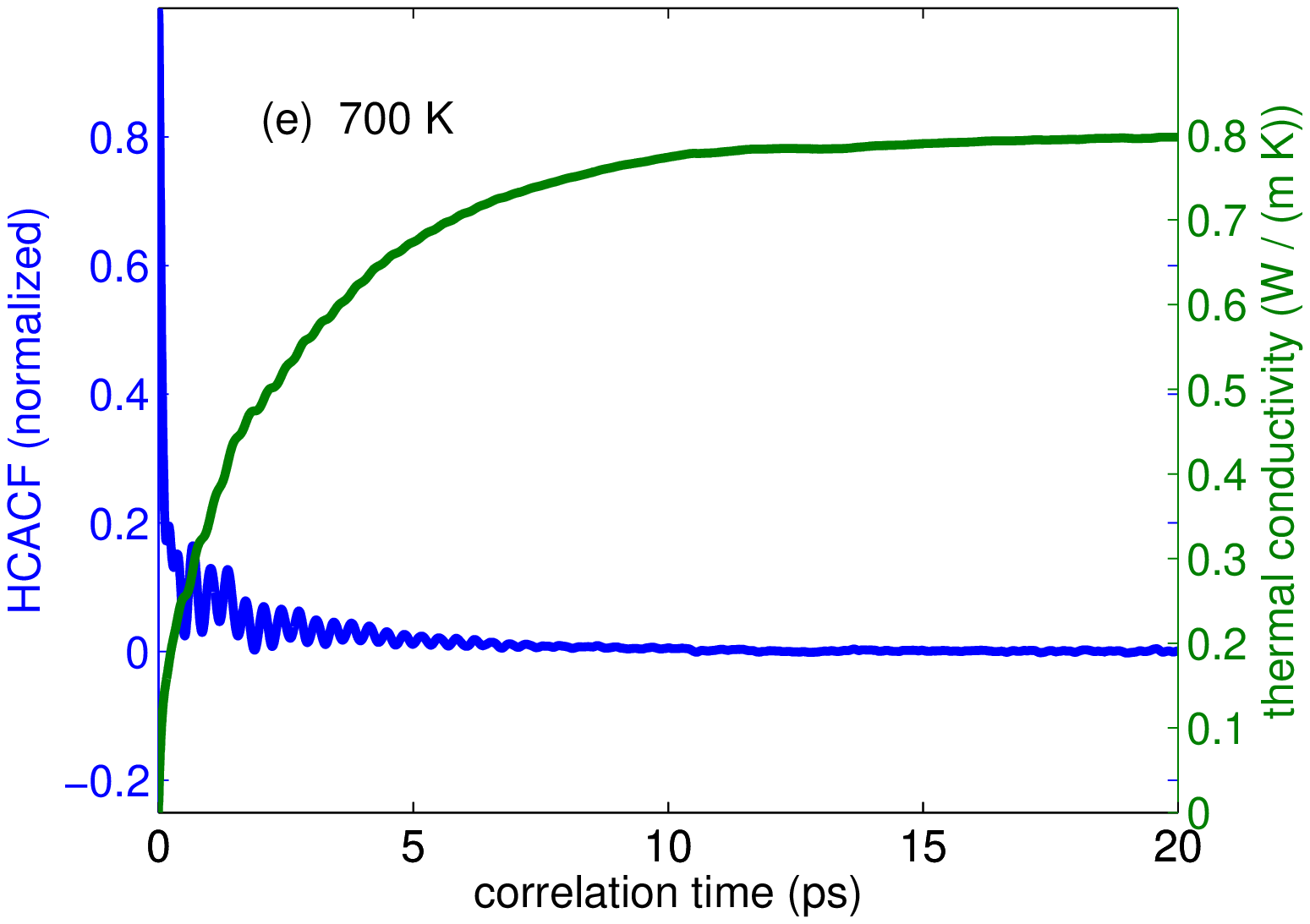}&
  \includegraphics[width=2.2 in]{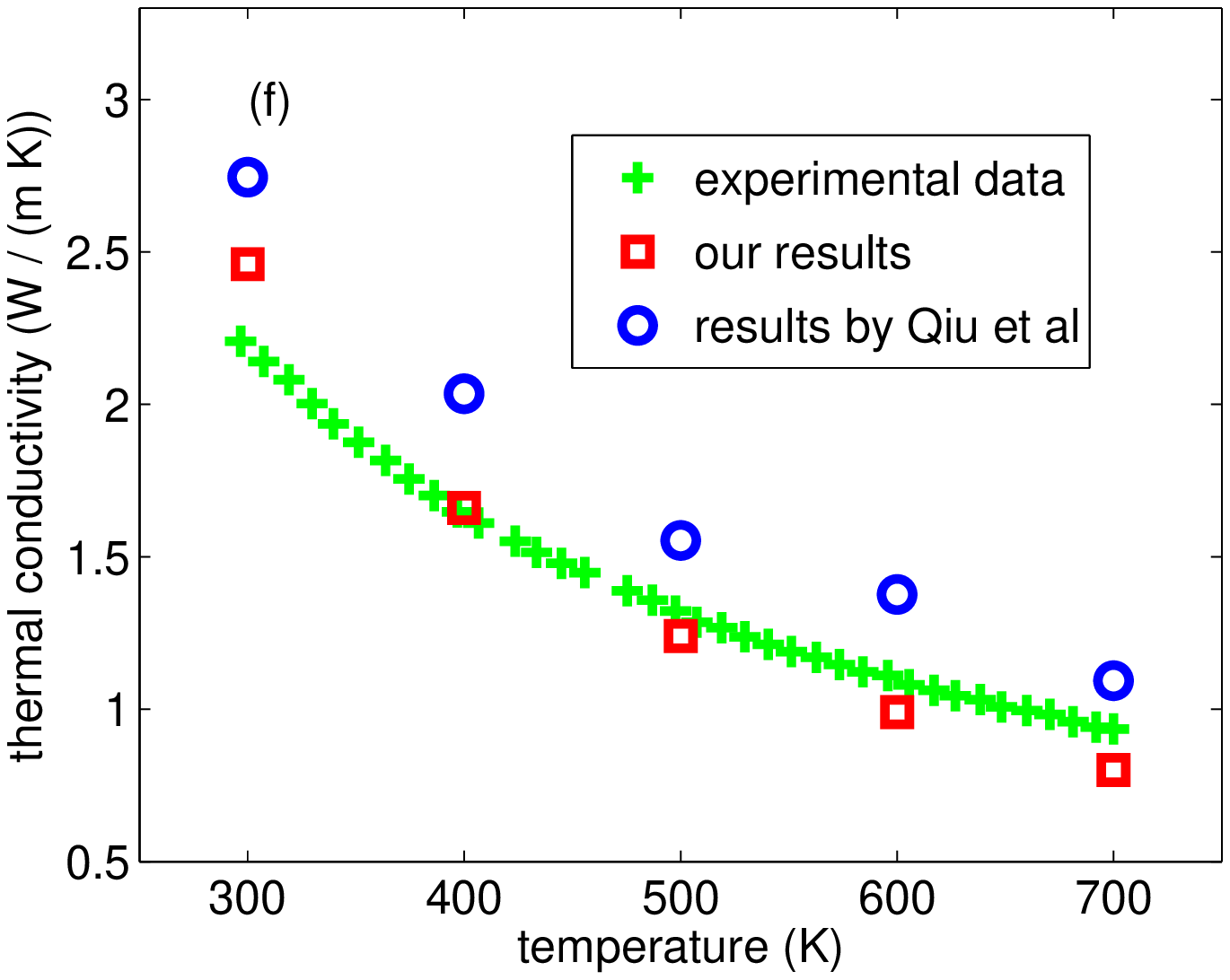}\\
\end{tabular}
  \caption{(Color online)
           HCACFs and RTCs (a-e) of PbTe as a function of the
           correlation time and the converged lattice thermal conductivity as a function
           of temperature (f). In (f), experimental data \cite{fedorov1969}
           and the MD results obtained by Qiu \textit{et al} \cite{qiu2011} 
           are also presented for comparison.}
  \label{figure:pbte_kappa}
\end{center}
\end{figure*}

The calculated HCACFs and RTCs at different temperatures
for PbTe using the GPU code in the block-scheme are presented in
Fig \ref{figure:pbte_kappa} (a-e). For PbTe, a number of
$10^7$ production steps is not sufficient to obtain smooth curves
for HCACF and RTC.
The curves presented in Fig \ref{figure:pbte_kappa} (a-e)
are obtained by setting the number of production
steps to be  $N_p = 10^8$. It can be seen that even if
there exists high frequency oscillations
(caused by optical phonons \cite{mcgaughey2004b}) in the HCACF,
the RTC still exhibits a very
smooth behavior as long as we collect
sufficiently many heat current data.

Figure \ref{figure:pbte_kappa} (f)  presents our calculated
lattice thermal conductivities of PbTe at different temperatures. 
We have tested the size effects and found that
a simulation size of $N = 4\times4\times4\times8=512$ is sufficient.
Our results agree well with the original results obtained by 
Qiu \textit{et al} \cite{qiu2011}. Both results agree fairly with
the experimental data \cite{fedorov1969}.

\subsection{A demonstration of the finite-size effect}

We stressed several times that the Green-Kubo method requires only 
small system sizes for well-converged results. 
Now, we give a demonstration of this important fact. 
Figure 8 presents the calculated thermal conductivities of solid argon   
at zero pressure and 10 K with different simulation sizes:
$N =$ 32, 108, 256, 500, 864, and 1372. 
We choose to demonstrate the finite-size effect
for the lowest temperature case, since this case gives the largest value
of thermal conductivity (and phonon mean free path), and a system with
larger thermal conductivity usually presents more prominent finite-size effect.
From Fig. 8 we see that, the finite-size effect can be eliminated even by
using a system of 256 argon atoms. The small finite-size effect is one
of the most advantages of the Green-Kubo method over the direct method,
and makes our new force evaluation scheme (the block-scheme) very useful.

\begin{figure}
\begin{center}
  \includegraphics[width=3 in]{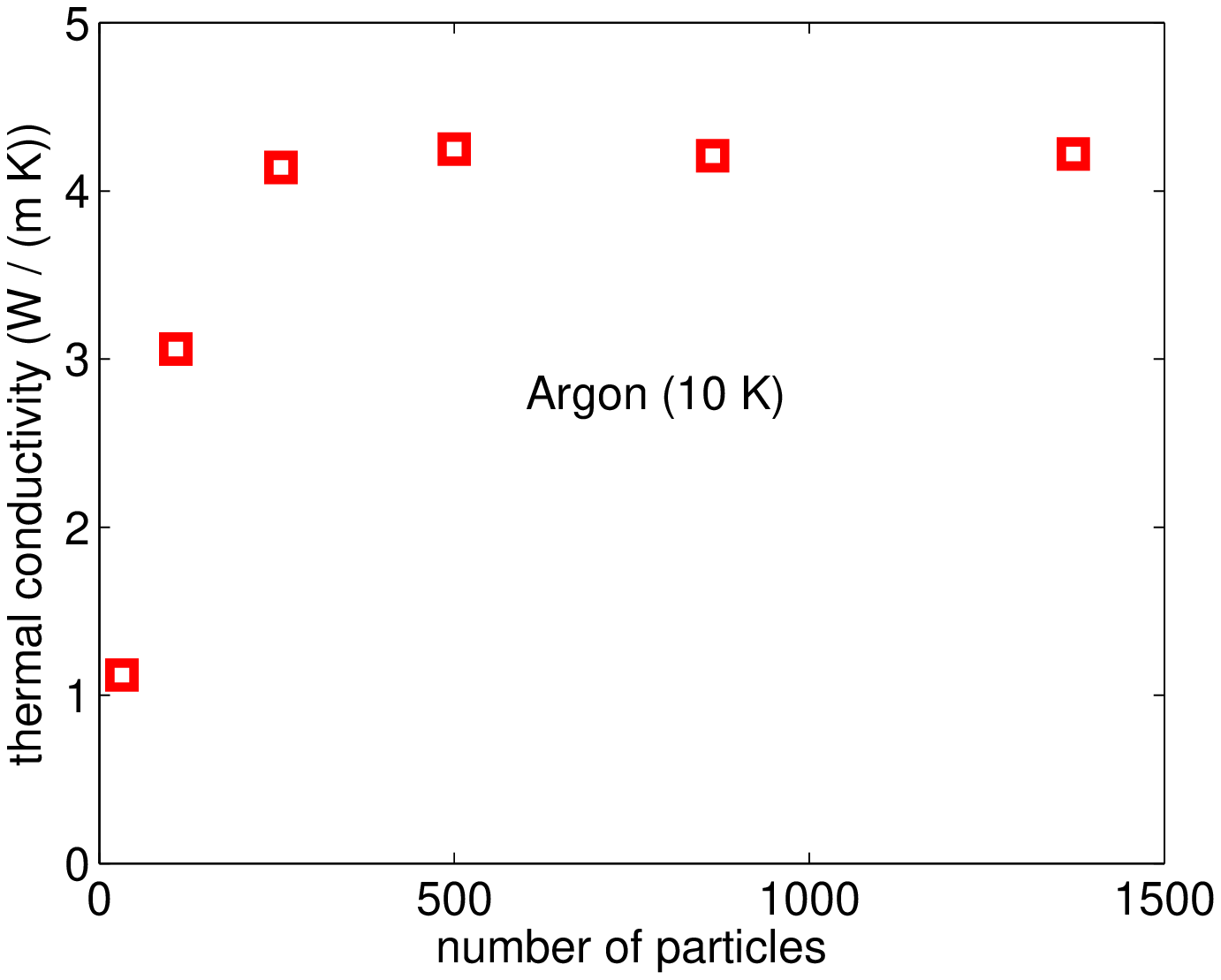}
  \caption{(Color online) Lattice thermal conductivity of solid argon at 0 Pa
           and 10 K as a function of the simulation size. }
  \label{figure:size_effect}
\end{center}
\end{figure}

\section{Conclusions}
\label{section:Conclusion}

In conclusion, we presented in detail the development and optimization
of a molecular dynamics
simulation program fully implemented in the GPU, which can calculate the lattice
thermal conductivity using the Green-Kubo formula. For the most time-consuming
part, the force evaluation part, we compared two alternative approaches,
a thread-scheme where the total force for a particle is accumulated in
a single thread and a block scheme where the pair forces for a particle
are distributively calculated in different threads within a block
and summed up using shared memory to obtain the total force of
the given particle. For both LJ and RI potentials,
the block-scheme outperforms the thread-scheme for smaller systems.
This makes the block-scheme particularly preferable for thermal conductivity
calculations using the Green-Kubo approach, which is more
demanding on the simulation time rather than the simulation size.
For large systems, the speedup factors obtained reach about one
hundred and three hundred for LJ and RI potentials, respectively.
The higher acceleration rate for the RI potential compared with 
that for the LJ potential results
from its higher arithmetic intensity, defined as the number of
arithmetic operations divided by the number of memory operations.
The correctness of our implementation is validated by calculating
the lattice thermal conductivities of solid argon and PbTe.

Both the LJ and the RI potentials considered in this work are very 
simple; they are pair-wise in nature. Generalization of our work to more 
general and complicated potentials deserves further consideration. 
It is interesting to consider 
the acceleration of the bond-order potential, which would
find interesting applications in carbon nanostructures. 
Since the bond-order potential also has high arithmetic intensity,
we expect that a well-designed GPU implementation of the bond-order potential 
can also lead to high acceleration rates.

In this work, we only applied our MD program to thermal conductivity calculations.
However, our program contains most of the essential parts of a general MD program.
Thus, one can modify it to study other problems. 
Our code is available upon request.

\section*{Acknowledgements}
This research has been supported by the Academy of Finland through
its Centres of Excellence Program (project no. 251748).

\bibliographystyle{model1-num-names}
\bibliography{zheyong}

\begin{thebibliography}{38}
\expandafter\ifx\csname natexlab\endcsname\relax\def\natexlab#1{#1}\fi
\providecommand{\bibinfo}[2]{#2}
\ifx\xfnm\relax \def\xfnm[#1]{\unskip,\space#1}\fi
\bibitem[{Belleman et~al.(2008)Belleman, Bedorf, and Zwart}]{belleman2008}
\bibinfo{author}{R.~G. Belleman}, \bibinfo{author}{J.~Bedorf},
  \bibinfo{author}{S.~F.~P. Zwart},
\newblock \bibinfo{title}{High performance direct gravitational ${N}$-body
  simulations on graphics processing units - {II}: an implementation in
  {CUDA}},
\newblock \bibinfo{journal}{New Astronomy} \bibinfo{volume}{13}
  (\bibinfo{year}{2008}) \bibinfo{pages}{103--112}.
\bibitem[{Yang et~al.(2007)Yang, Wang, and Chen}]{yang2007}
\bibinfo{author}{J.~Yang}, \bibinfo{author}{Y.~Wang},
  \bibinfo{author}{Y.~Chen},
\newblock \bibinfo{title}{{GPU} accelerated molecular dynamics simulation of
  thermal conductivities},
\newblock \bibinfo{journal}{J. Comp. Phys.} \bibinfo{volume}{221}
  (\bibinfo{year}{2007}) \bibinfo{pages}{799--804}.
\bibitem[{Stone et~al.(2007)Stone, Phillips, Freddolino, Hardy, Trabuco, and
  Schulten}]{stone2007}
\bibinfo{author}{J.~E. Stone}, \bibinfo{author}{J.~C. Phillips},
  \bibinfo{author}{P.~L. Freddolino}, \bibinfo{author}{D.~J. Hardy},
  \bibinfo{author}{L.~G. Trabuco}, \bibinfo{author}{K.~Schulten},
\newblock \bibinfo{title}{Accelerating molecular modeling applications with
  graphics processors},
\newblock \bibinfo{journal}{J. Comp. Chem.} \bibinfo{volume}{28}
  (\bibinfo{year}{2007}) \bibinfo{pages}{72618--2640}.
\bibitem[{van Meel et~al.(2008)van Meel, Arnold, Frenkel, {Portegies Zwart},
  and Belleman}]{vanmeel2008}
\bibinfo{author}{J.~A. van Meel}, \bibinfo{author}{A.~Arnold},
  \bibinfo{author}{D.~Frenkel}, \bibinfo{author}{S.~F. {Portegies Zwart}},
  \bibinfo{author}{R.~G. Belleman},
\newblock \bibinfo{title}{Harvesting graphics power for {MD} simulations},
\newblock \bibinfo{journal}{Molecular Simulation} \bibinfo{volume}{34}
  (\bibinfo{year}{2008}) \bibinfo{pages}{259--266}.
\bibitem[{Anderson et~al.(2008)Anderson, Lorenz, and Travesset}]{anderson2008}
\bibinfo{author}{J.~A. Anderson}, \bibinfo{author}{C.~D. Lorenz},
  \bibinfo{author}{A.~Travesset},
\newblock \bibinfo{title}{General purpose molecular dynamics simulations fully
  implemented on graphics processing units},
\newblock \bibinfo{journal}{J. Comp. Phys.} \bibinfo{volume}{227}
  (\bibinfo{year}{2008}) \bibinfo{pages}{5342--5359}.
\bibitem[{Liu et~al.(2008)Liu, Schmidt, Voss, and M{\"u}ller-Wittig}]{liu2008}
\bibinfo{author}{W.~Liu}, \bibinfo{author}{B.~Schmidt},
  \bibinfo{author}{G.~Voss}, \bibinfo{author}{W.~M{\"u}ller-Wittig},
\newblock \bibinfo{title}{Accelerating molecular dynamics simulations using
  graphics processing units with {CUDA}},
\newblock \bibinfo{journal}{Comp. Phys. Commun.} \bibinfo{volume}{179}
  (\bibinfo{year}{2008}) \bibinfo{pages}{634--641}.
\bibitem[{Friedrichs et~al.(2009)Friedrichs, Eastman, Vaidyanathan, Houston,
  Legrand, Beberg, Ensign, Bruns, and Pande}]{friedrichs2009}
\bibinfo{author}{M.~S. Friedrichs}, \bibinfo{author}{P.~Eastman},
  \bibinfo{author}{V.~Vaidyanathan}, \bibinfo{author}{M.~Houston},
  \bibinfo{author}{S.~Legrand}, \bibinfo{author}{A.~L. Beberg},
  \bibinfo{author}{D.~L. Ensign}, \bibinfo{author}{C.~M. Bruns},
  \bibinfo{author}{V.~S. Pande},
\newblock \bibinfo{title}{Accelerating molecular dynamic simulation on graphics
  processing units},
\newblock \bibinfo{journal}{J Comput Chem.} \bibinfo{volume}{30}
  (\bibinfo{year}{2009}) \bibinfo{pages}{864--872}.
\bibitem[{Rapaport(2011)}]{rapaport2011}
\bibinfo{author}{D.~C. Rapaport},
\newblock \bibinfo{title}{Enhanced molecular dynamics performance with a
  programmable graphics processor},
\newblock \bibinfo{journal}{Comp. Phys. Commun.} \bibinfo{volume}{182}
  (\bibinfo{year}{2011}) \bibinfo{pages}{926--934}.
\bibitem[{Preis et~al.(2009)Preis, Virnau, Paul, and Schneider}]{preis2009}
\bibinfo{author}{T.~Preis}, \bibinfo{author}{P.~Virnau},
  \bibinfo{author}{W.~Paul}, \bibinfo{author}{J.~J. Schneider},
\newblock \bibinfo{title}{{GPU} accelerated {M}onte {C}arlo simulation of the
  {2D} and {3D} {I}sing model},
\newblock \bibinfo{journal}{J. Comp. Phys.} \bibinfo{volume}{228}
  (\bibinfo{year}{2009}) \bibinfo{pages}{4468--4477}.
\bibitem[{Anderson et~al.(2007)Anderson, Goddard, and
  Schr{\"o}der}]{anderson2007}
\bibinfo{author}{A.~G. Anderson}, \bibinfo{author}{W.~A. Goddard},
  \bibinfo{author}{P.~Schr{\"o}der},
\newblock \bibinfo{title}{Quantum {M}onte {C}arlo on graphical processing
  units},
\newblock \bibinfo{journal}{Comp. Phys. Commun.} \bibinfo{volume}{177}
  (\bibinfo{year}{2007}) \bibinfo{pages}{298--306}.
\bibitem[{Ihnatsenka(2012)}]{ihnatsenka2012}
\bibinfo{author}{S.~Ihnatsenka},
\newblock \bibinfo{title}{Computation of electron quantum transport in graphene
  nanoribbons using {GPU}},
\newblock \bibinfo{journal}{Comp. Phys. Commun.} \bibinfo{volume}{183}
  (\bibinfo{year}{2012}) \bibinfo{pages}{543--546}.
\bibitem[{Siro and Harju(2012)}]{siro2012}
\bibinfo{author}{T.~Siro}, \bibinfo{author}{A.~Harju},
\newblock \bibinfo{title}{Exact diagonalization of the {H}ubbard model on
  graphics processing units},
\newblock \bibinfo{journal}{Comp. Phys. Commun.} \bibinfo{volume}{183}
  (\bibinfo{year}{2012}) \bibinfo{pages}{1884--1889}.
\bibitem[{Yao et~al.(2005)Yao, Wang, Li, and Liu}]{yao2005}
\bibinfo{author}{Z.~Yao}, \bibinfo{author}{J.~S. Wang},
  \bibinfo{author}{B.~Li}, \bibinfo{author}{G.~R. Liu},
\newblock \bibinfo{title}{Thermal conduction of carbon nanotubes using
  molecular dynamics},
\newblock \bibinfo{journal}{Phys. Rev. B} \bibinfo{volume}{71}
  (\bibinfo{year}{2005}) \bibinfo{pages}{085417}.
\bibitem[{Schelling et~al.(2002)Schelling, Phillpot, and
  Keblinski}]{schelling2002}
\bibinfo{author}{P.~K. Schelling}, \bibinfo{author}{S.~R. Phillpot},
  \bibinfo{author}{P.~Keblinski},
\newblock \bibinfo{title}{Comparison of atomic-level simulation methods for
  computing thermal conductivity},
\newblock \bibinfo{journal}{Phys. Rev. B} \bibinfo{volume}{65}
  (\bibinfo{year}{2002}) \bibinfo{pages}{144306}.
\bibitem[{Sellan et~al.(2010)Sellan, Landry, Turney, McGaughey, and
  Amon}]{sellan2010}
\bibinfo{author}{D.~P. Sellan}, \bibinfo{author}{E.~S. Landry},
  \bibinfo{author}{J.~E. Turney}, \bibinfo{author}{A.~J.~H. McGaughey},
  \bibinfo{author}{C.~H. Amon},
\newblock \bibinfo{title}{Size effects in molecular dynamics thermal
  conductivity predictions},
\newblock \bibinfo{journal}{Phys. Rev. B} \bibinfo{volume}{81}
  (\bibinfo{year}{2010}) \bibinfo{pages}{214305}.
\bibitem[{Green(1954)}]{green1954}
\bibinfo{author}{M.~S. Green},
\newblock \bibinfo{title}{Markoff random processes and the statistical
  mechanics of time-dependent phenomena. {II}. irreversible processes in
  fluids},
\newblock \bibinfo{journal}{J. Chem. Phys.} \bibinfo{volume}{22}
  (\bibinfo{year}{1954}) \bibinfo{pages}{398}.
\bibitem[{Kubo(1957)}]{kubo1957}
\bibinfo{author}{R.~Kubo},
\newblock \bibinfo{title}{Statistical-mechanical theory of irreversible
  processes. {I}. general theory and simple applications to magnetic and
  conduction problems},
\newblock \bibinfo{journal}{J. Phys. Soc. Jpn.} \bibinfo{volume}{12}
  (\bibinfo{year}{1957}) \bibinfo{pages}{570}.
\bibitem[{McQuarrie(2000)}]{mcquarrie2000}
\bibinfo{author}{D.~A. McQuarrie}, \bibinfo{title}{Statistical Mechanics},
  \bibinfo{publisher}{University Science Books, Sausalito},
  \bibinfo{year}{2000}.
\bibitem[{Kaburaki et~al.(1999)Kaburaki, Li, and Yip}]{kaburaki1999}
\bibinfo{author}{H.~Kaburaki}, \bibinfo{author}{J.~Li},
  \bibinfo{author}{S.~Yip},
\newblock \bibinfo{title}{Thermal conductivity of solid argon by molecular
  dynamics},
\newblock \bibinfo{journal}{Mater. Res. Soc. Symp. Proc.} \bibinfo{volume}{538}
  (\bibinfo{year}{1999}) \bibinfo{pages}{503}.
\bibitem[{McGaughey and Kaviany(2004)}]{mcgaughey2004}
\bibinfo{author}{A.~J.~H. McGaughey}, \bibinfo{author}{M.~Kaviany},
\newblock \bibinfo{title}{Thermal conductivity decomposition and analysis using
  molecular dynamics simulations},
\newblock \bibinfo{journal}{Int. J. Heat Mass Transfer} \bibinfo{volume}{47}
  (\bibinfo{year}{2004}) \bibinfo{pages}{1783}.
\bibitem[{Tretiakova and Scandolo(2004)}]{tretiakova2004}
\bibinfo{author}{K.~V. Tretiakova}, \bibinfo{author}{S.~Scandolo},
\newblock \bibinfo{title}{Thermal conductivity of solid argon from molecular
  dynamics simulations},
\newblock \bibinfo{journal}{J. Chem. Phys.} \bibinfo{volume}{120}
  (\bibinfo{year}{2004}) \bibinfo{pages}{3765}.
\bibitem[{Chen et~al.(2004)Chen, Lukes, Li, Yang, and Wu}]{chen2004}
\bibinfo{author}{Y.~Chen}, \bibinfo{author}{J.~R. Lukes},
  \bibinfo{author}{D.~Li}, \bibinfo{author}{J.~Yang}, \bibinfo{author}{Y.~Wu},
\newblock \bibinfo{title}{Thermal expansion and impurity effects on lattice
  thermal conductivity of solid argon},
\newblock \bibinfo{journal}{J. Chem. Phys.} \bibinfo{volume}{120}
  (\bibinfo{year}{2004}) \bibinfo{pages}{3841}.
\bibitem[{Kaburaki et~al.(2007)Kaburaki, Li, Yip, and Kimizuka}]{kaburaki2007}
\bibinfo{author}{H.~Kaburaki}, \bibinfo{author}{J.~Li},
  \bibinfo{author}{S.~Yip}, \bibinfo{author}{H.~Kimizuka},
\newblock \bibinfo{title}{Dynamical thermal conductivity of argon crystal},
\newblock \bibinfo{journal}{J. Appl. Phys.} \bibinfo{volume}{102}
  (\bibinfo{year}{2007}) \bibinfo{pages}{043514}.
\bibitem[{Chen et~al.(2005)Chen, Li, Lukes, Ni, and Chen}]{chen2005}
\bibinfo{author}{Y.~Chen}, \bibinfo{author}{D.~Li}, \bibinfo{author}{J.~R.
  Lukes}, \bibinfo{author}{Z.~Ni}, \bibinfo{author}{M.~Chen},
\newblock \bibinfo{title}{Minimum superlattice thermal conductivity from
  molecular dynamics},
\newblock \bibinfo{journal}{Phys. Rev. B} \bibinfo{volume}{72}
  (\bibinfo{year}{2005}) \bibinfo{pages}{174302}.
\bibitem[{Landry et~al.(2008)Landry, Hussein, and McGaughey}]{landry2008}
\bibinfo{author}{E.~S. Landry}, \bibinfo{author}{M.~I. Hussein},
  \bibinfo{author}{A.~J.~H. McGaughey},
\newblock \bibinfo{title}{Complex superlattice unit cell designs for reduced
  thermal conductivity},
\newblock \bibinfo{journal}{Phys. Rev. B} \bibinfo{volume}{77}
  (\bibinfo{year}{2008}) \bibinfo{pages}{0184302}.
\bibitem[{Wolf et~al.(1999)Wolf, Keblinski, Phillpot, and
  Eggebrecht}]{wolf1999}
\bibinfo{author}{D.~Wolf}, \bibinfo{author}{P.~Keblinski},
  \bibinfo{author}{S.~R. Phillpot}, \bibinfo{author}{J.~Eggebrecht},
\newblock \bibinfo{title}{Exact method for the simulation of coulombic systems
  by spherically truncated, pairwise $r^{-1}$ summation},
\newblock \bibinfo{journal}{J. Chem. Phys.} \bibinfo{volume}{110}
  (\bibinfo{year}{1999}) \bibinfo{pages}{8255}.
\bibitem[{Fennell and Gezelter(2006)}]{fennell2006}
\bibinfo{author}{C.~J. Fennell}, \bibinfo{author}{J.~D. Gezelter},
\newblock \bibinfo{title}{Is the {E}wald summation still necessary? pairwise
  alternatives to the accepted standard for long-range electrostatics},
\newblock \bibinfo{journal}{J. Chem. Phys.} \bibinfo{volume}{124}
  (\bibinfo{year}{2006}) \bibinfo{pages}{234104}.
\bibitem[{McGaughey and Kaviany(2004)}]{mcgaughey2004b}
\bibinfo{author}{A.~J.~H. McGaughey}, \bibinfo{author}{M.~Kaviany},
\newblock \bibinfo{title}{Thermal conductivity decomposition and analysis using
  molecular dynamics simulations},
\newblock \bibinfo{journal}{Int. J. Heat Mass Transfer} \bibinfo{volume}{47}
  (\bibinfo{year}{2004}) \bibinfo{pages}{1799--1816}.
\bibitem[{Kulkarni and Zhou(2006)}]{kulkarni2006}
\bibinfo{author}{A.~J. Kulkarni}, \bibinfo{author}{M.~Zhou},
\newblock \bibinfo{title}{Size-dependent thermal conductivity of zinc oxide
  nanobelts},
\newblock \bibinfo{journal}{Appl. Phys. Lett.} \bibinfo{volume}{88}
  (\bibinfo{year}{2006}) \bibinfo{pages}{141921}.
\bibitem[{Qiu et~al.(2012)Qiu, Bao, Zhang, Wu, and Ruan}]{qiu2011}
\bibinfo{author}{B.~Qiu}, \bibinfo{author}{H.~Bao}, \bibinfo{author}{G.~Zhang},
  \bibinfo{author}{Y.~Wu}, \bibinfo{author}{X.~Ruan},
\newblock \bibinfo{title}{Molecular dynamics simulations of lattice thermal
  conductivity and spectral phonon mean free path of {P}b{T}e: Bulk and
  nanostructures},
\newblock \bibinfo{journal}{Comp. Mater. Sci.} \bibinfo{volume}{53}
  (\bibinfo{year}{2012}) \bibinfo{pages}{278--285}.
\bibitem[{Qiu et~al.(2008)Qiu, Bao, and Ruan}]{qiu2008}
\bibinfo{author}{B.~Qiu}, \bibinfo{author}{H.~Bao}, \bibinfo{author}{X.~Ruan},
\newblock \bibinfo{title}{Multiscale simulations of thermoelectric properties
  of {P}b{Te}},
\newblock \bibinfo{journal}{ASME 2008 3rd Energy Nanotechnology International
  COnference}  (\bibinfo{year}{2008}) \bibinfo{pages}{45}.
\bibitem[{Che et~al.(2000)Che, Cagin, Deng, and Goddard}]{che2000}
\bibinfo{author}{J.~Che}, \bibinfo{author}{T.~Cagin},
  \bibinfo{author}{W.~Deng}, \bibinfo{author}{W.~A. Goddard},
\newblock \bibinfo{title}{Thermal conductivity of diamond and related materials
  from molecular dynamics simulations},
\newblock \bibinfo{journal}{J. Chem. Phys.} \bibinfo{volume}{113}
  (\bibinfo{year}{2000}) \bibinfo{pages}{6888}.
\bibitem[{Henry and Chen(2008)}]{henry2008}
\bibinfo{author}{A.~Henry}, \bibinfo{author}{G.~Chen},
\newblock \bibinfo{title}{High thermal conductivity of single polyethylene
  chains using molecular dynamics simulations},
\newblock \bibinfo{journal}{Phys. Rev. Lett.} \bibinfo{volume}{101}
  (\bibinfo{year}{2008}) \bibinfo{pages}{235502}.
\bibitem[{NVIDIA(2011)}]{cuda2011}
\bibinfo{author}{NVIDIA},
\newblock \bibinfo{title}{{CUDA C} programming guide, version 4.0}
  (\bibinfo{year}{2011}).
\bibitem[{Verlet(1967)}]{verlet1967}
\bibinfo{author}{L.~Verlet},
\newblock \bibinfo{title}{Computer `experiments' on classical fluids. {I}.
  thermodynamical properties of {L}ennard-{J}ones molecules},
\newblock \bibinfo{journal}{Phys. Rev.} \bibinfo{volume}{159}
  (\bibinfo{year}{1967}) \bibinfo{pages}{98--103}.
\bibitem[{Ashcroft and Mermin(1976)}]{ashcroft1976}
\bibinfo{author}{N.~W. Ashcroft}, \bibinfo{author}{N.~D. Mermin},
  \bibinfo{title}{Solid State Physics}, \bibinfo{publisher}{Saunders College,
  Orlando, FL}, \bibinfo{year}{1976}.
\bibitem[{Houston et~al.(1968)Houston, Strakna, and Belson}]{houston1968}
\bibinfo{author}{B.~Houston}, \bibinfo{author}{R.~Strakna},
  \bibinfo{author}{H.~Belson},
\newblock \bibinfo{title}{Elastic constants, thermal expansion, and {D}ebye
  temperature of lead telluride},
\newblock \bibinfo{journal}{J. Appl. Phys.} \bibinfo{volume}{39}
  (\bibinfo{year}{1968}) \bibinfo{pages}{3913--3916}.
\bibitem[{Fedorov and Machuev(1969)}]{fedorov1969}
\bibinfo{author}{V.~I. Fedorov}, \bibinfo{author}{V.~I. Machuev},
\newblock \bibinfo{title}{The thermal conductivity of {P}b{T}e, {S}n{T}e, and
  {G}e{T}e in the solid and liquid phases},
\newblock \bibinfo{journal}{Sov. Phys. Solid State, USSR} \bibinfo{volume}{11}
  (\bibinfo{year}{1969}) \bibinfo{pages}{1116}.

\end{thebibliography}

\end{document}